\documentclass[%
 aip,
 amsmath,amssymb,
 reprint,%
]{revtex4-1}

\usepackage{graphicx}
\usepackage{dcolumn}
\usepackage{bm}

\usepackage[utf8]{inputenc}
\usepackage[T1]{fontenc}
\usepackage{mathptmx}
\usepackage{etoolbox}
\usepackage{float}
\makeatletter
\def\@email#1#2{%
 \endgroup
 \patchcmd{\titleblock@produce}
  {\frontmatter@RRAPformat}
  {\frontmatter@RRAPformat{\produce@RRAP{*#1\href{mailto:#2}{#2}}}\frontmatter@RRAPformat}
  {}{}
}%
\usepackage[mathscr]{euscript}
\usepackage{color}
\usepackage{xcolor,soul}
\usepackage{xspace}
\usepackage{amsmath}
\usepackage[bb=boondox]{mathalfa}
\usepackage{mathrsfs}
\usepackage{multirow}

\definecolor{darkgreen}{rgb}{0,0.5,0}

\newcommand{\uL}{\ensuremath{\,\mu\mathrm{L}}\xspace}

\DeclareMathAlphabet{\mymathbb}{U}{BOONDOX-ds}{m}{n}

\renewcommand{\deg}{\ensuremath{{}^{\circ}}\xspace}

\newcommand{\abs}[1]{|#1|}

\def\Cth{\ensuremath{{}^{13}\mathrm{C}}\xspace}

\newcommand{\Ttwo}{T_2}

\newcommand{\wnut}{\omega_{\rm nut}}

\newcommand{\Rh}{$\mathrm{^{103}Rh}$\xspace}
\newcommand{\Proton}{$\mathrm{^{1}H}$\xspace}
\newcommand{\taurelax}{\tau_\mathrm{relax}}
\include{References/RefSets}
\renewcommand\hl[1]{{#1}}
\makeatother
\begin{document}

\preprint{AIP/123-QED}

\title{The \Rh NMR Spectroscopy and Relaxometry of the Rhodium Formate Paddlewheel Complex}
\author{Harry Harbor Collins}
\affiliation{School of Chemistry, University of Southampton, SO17 1BJ, UK}

\author{Mohamed Sabba}
\affiliation{School of Chemistry, University of Southampton, SO17 1BJ, UK}

\author{Gamal Moustafa}
\affiliation{School of Chemistry, University of Southampton, SO17 1BJ, UK}

\author{Bonifac Legrady}
\affiliation{School of Chemistry, University of Southampton, SO17 1BJ, UK}

\author{Murari Soundararajan}
\affiliation{School of Chemistry, University of Southampton, SO17 1BJ, UK}

\author{Markus Leutzsch}
\affiliation{Max-Planck-Institut für Kohlenforschung, Kaiser-Wilhelm-Platz 1, Mülheim an der Ruhr, 45470, Germany}

\author{Malcolm H. Levitt}
 \email{mhl@soton.ac.uk}
 \affiliation{School of Chemistry, University of Southampton, SO17 1BJ, UK}

 \homepage{http://www.Second.institution.edu/~Charlie.Author.}

\date{\today}

\begin{abstract}
The NMR spectroscopy of spin-1/2 nuclei with low gyromagnetic ratio is challenging, due to the low NMR signal strength.  Methodology for the rapid acquisition of \Rh NMR parameters is demonstrated for the case of the rhodium formate ``paddlewheel" complex $\mathrm{Rh_2(HCO_2)_4}$. A scheme is described for enhancing the \Rh signal strength by polarization transfer from \Proton nuclei and which also greatly reduces the interference from ringing artifacts, a common hurdle for the direct observation of low-$\gamma$ nuclei. The \Rh relaxation time constants $T_1$ and $T_2$ are \hl{measured within 20 minutes by using} $^{1}$H-detected experiments. The field-dependence of the \Rh $T_1$ is measured. The high-field relaxation is dominated by the chemical shift anisotropy (CSA) mechanism. The \Rh shielding anisotropy is found to be very large: $|\Delta\sigma|=9900\pm540\mathrm{\,ppm}$. This estimate is compared with \hl{density functional theory} calculations. 
\end{abstract}

\maketitle

\section{\label{sec:Intro}Introduction}

\begin{figure}[tbh]
\hspace*{-0.68cm}
\centering
\includegraphics[trim={4cm 1cm 4cm 1cm},clip,width=1\columnwidth]{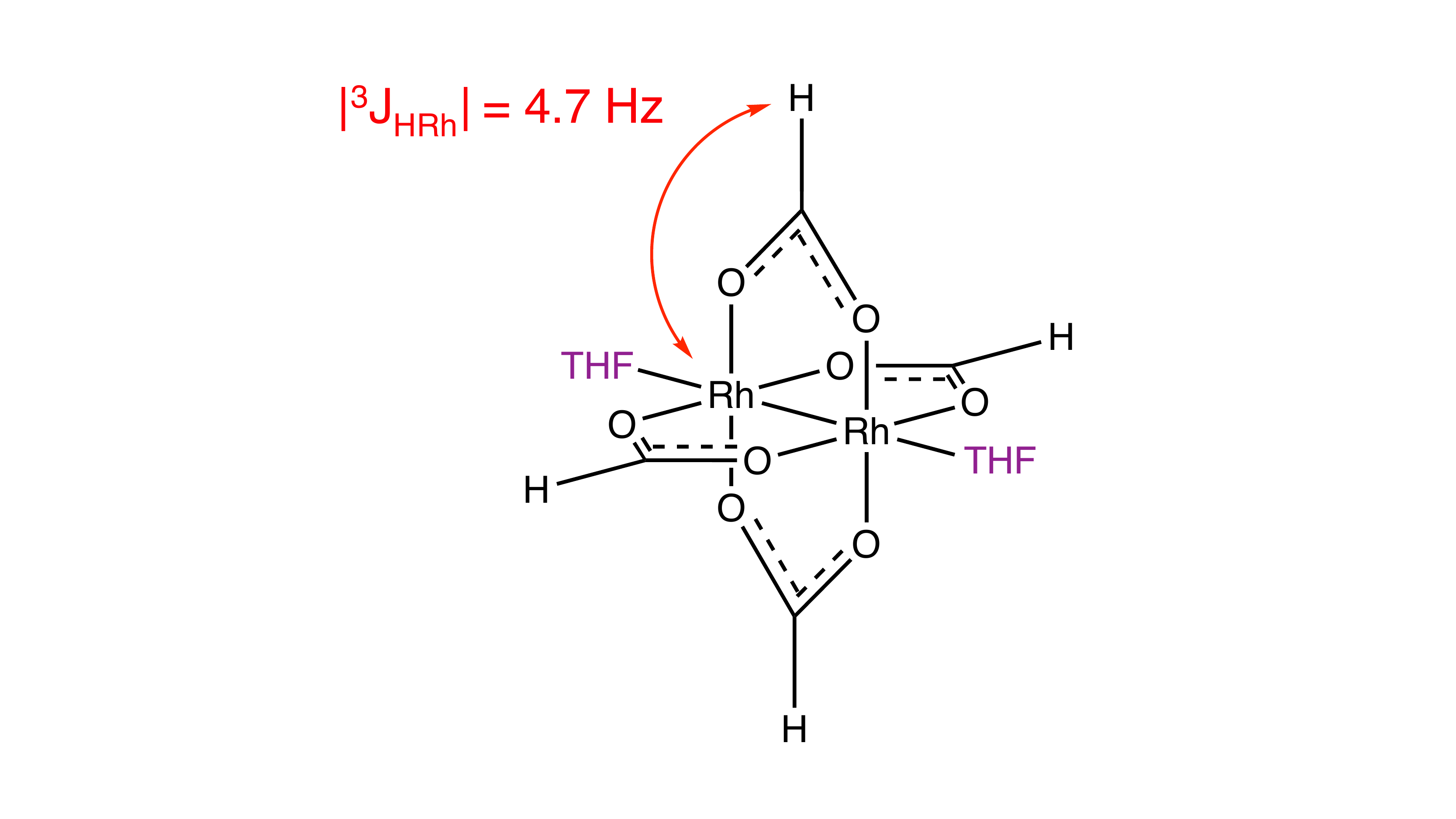}
\setlength{\belowcaptionskip}{-0pt}
\caption{\label{fig:Rhformatestruc} 
Molecular structure of the rhodium formate paddlewheel complex ligated by solvent tetrahydrofuran (THF) molecules at the axial sites. This work exploits the $^3J_\mathrm{RhH}$ scalar couplings for polarisation transfer  between the $^{103}$Rh and $^1$H nuclei.
}
\end{figure}

Rhodium paddlewheel complexes have attracted significant attention due to their unique properties and diverse applications where they have played roles as catalysts and potential anticancer agents.\cite{erck_studies_1974,fandzloch_first_2020,ohata_rhodium_2018,lin_dirhodium_2021,hrdina_dirhodiumiiii_2021} 
These complexes consist of two rhodium atoms bridged by four carboxylate ligands, forming a lantern-like structure, with some resemblance to the paddlewheels of a river boat. A typical example is rhodium formate, $\mathrm{Rh_2(HCO_2)_4}$, see figure~\ref{fig:Rhformatestruc}.

Nuclear magnetic resonance (NMR) is a powerful probe of the properties of rhodium complexes. \Rh carries the distinction of being one of only 4 (with $^{19}$F, $^{31}$P, and $^{89}$Y) spin-1/2 nuclei with a natural abundance of 100\%. Nevertheless, it has been relatively neglected by spectroscopists: \Rh is a member of what Mann dubbed "the Cinderella nuclei"\cite{mann_cinderella_1991} - transition metals with spin-1/2 but very low
magnetogyric ratio $\gamma$. The NMR of \Rh is associated with multiple experimental challenges leading to a relative scarcity of experimental data. However, many of these challenges have been successfully overcome by the creative application of modern NMR methodology, such as heteronuclear multiple-quantum (HMQC) NMR~\cite{calo_triple_2021}. However, although HMQC experiments allow the rapid acquisition of \Rh NMR spectra in suitable cases, it is not possible to estimate \Rh spin-lattice and spin-spin relaxation time constants through HMQC experiments. For this purpose, experiments exploiting \Rh magnetization are needed.

In this work, we utilise a variant of the PulsePol polarisation transfer technique~\cite{schwartz_robust_2018,tratzmiller_pulsed_2021,sabba_symmetry-based_2022} to enhance the \Rh NMR spectroscopy of the rhodium formate paddlewheel complex in solution.
We report (i) NMR methodology for the acquisition of directly detected \Rh spectra with effective ringing filtration; (ii) NMR methodology for the rapid measurement of \Rh $T_1$ and $T_2$ relaxation time constants over a range of magnetic field strengths. We observe a strong field dependence of the \Rh $T_1$, which is qualitatively consistent with a dominant chemical shift anisotropy relaxation mechanism. We estimate the \Rh shielding anisotropy by using information from \Cth and \Rh relaxation experiments in solution, and from \Cth solid-state NMR.

\section{\label{sec:Experimental}Experimental}

\subsection{Sample}

Experiments were performed on a saturated \hl{($\sim$10 mM)} solution of rhodium formate ($\mathrm{Rh_2(HCO_2)_4}$) dissolved in 500~\uL deuterated tetrahydrofuran (THF-d$_8$) contained in a Wilmad LPV 5~mL sample tube. The rhodium formate was synthesised from rhodium chloride using a reported procedure\cite{rempel_tetrakisacetatodirhodiumii_1972} and dried extensively under heated vacuum. The resulting rhodium formate solid was green in colour and dissolved in THF to produce a green solution.

\subsection{Solution NMR}

\Proton and \Rh Spectra were acquired at a magnetic field strength of 9.4~T using a standard commercial Bruker 5~mm NMR BBO probe (\Proton/$^2$H/$^{109}$Ag-$^{31}$P) equipped with a z-gradient with a maximum strength of 50 G cm$^{-1}$.

Proton resonances are referenced to the absolute frequency 400.14300 MHz; whereas \Rh resonances are referenced to an absolute frequency that is proportional to the protons ($\Xi\left(^{103}\rm{Rh}\right)$ = 3.16$\%$) per the most common convention~\cite{carlton_chapter_2008}.

Although the probe could be tuned to \Rh beyond the manufacturer specifications, it was set to mismatched (overcoupled) conditions to reduce ringdown times~\cite{chingas_overcoupling_1983,buess_acoustic_1978,gerothanassis_methods_1987,fukushima_spurious_1979}. The radiofrequency amplitudes on the $^{1}$H and \Rh channels were both adjusted to give an intentionally matched nutation frequency of 
$\wnut/(2\pi)\simeq 4\mathrm{\,kHz}$, corresponding to a 90\deg pulse duration of $62.5 \,\mu\mathrm{s}$. 

Additional isolation of the rf channels by electronic filters was found to be necessary 
- without the filters, noise on the \Rh channel was significant enough to preclude observation of other nuclei. At the preamplifier output we installed: a 30 MHz lowpass filter (Chemagnetics) on the \Rh channel, a 400 MHz bandpass filter (K\&L Microwave) on the $^{1}$H channel, and a 61 MHz bandpass filter (FSY Microwave) on the $^{2}$H lock channel.

To measure relaxation times as a function of magnetic field, the experiments used rapid sample shuttling from inside the 9.4~T magnet bore to regions of lower field outside the magnet bore. The shuttling was performed using a motorised fast shuttling system based on the design by Kiryutin~\cite{kiryutin_fast_2016}. The shuttling time was kept constant at 1 second.

The pulse sequences described below use the following elements:

\subsubsection{Composite pulses}

Composite pulses were used to minimize the effects of rf field inhomogeneity and are denoted by shaded black rectangles in the pulse sequence diagrams. 
All composite pulses are implemented using the symmetrized BB1 composite pulse scheme~\cite{wimperis_broadband_1994,cummins_tackling_2003} in which a simple pulse $\beta_{\phi}$ (where $\beta$ is the flip angle and $\phi$ is the phase) is replaced by:
\begin{equation}
(\beta/2)_{\phi}180_{\phi+\theta_W}360_{\phi+3\theta_W}180_{\phi+\theta_W}(\beta/2)_{\phi}
\end{equation}
Where $\theta_{W} = \arccos{(-\beta/(4\pi))}$. For the $\pi/2$ and $\pi$ flip angles used in this paper, this corresponds to the following sequences:
\begin{equation}\label{eq:composite90}
90_{\phi} \rightarrow 45 _{\phi}180_{\phi+97.18}360_{\phi+291.54}180_{\phi+97.18}45_{\phi}
\end{equation}
\begin{equation}\label{eq:composite180}
180_{\phi} \rightarrow 90_{\phi}180_{\phi+104.48}360_{\phi+313.43}180_{\phi+104.48}90_{\phi}
\end{equation}

\begin{figure}[tbh]
\hspace*{-0.68cm}
\centering
\includegraphics[trim={0cm 0cm 0cm 0cm},clip,width=1\columnwidth]{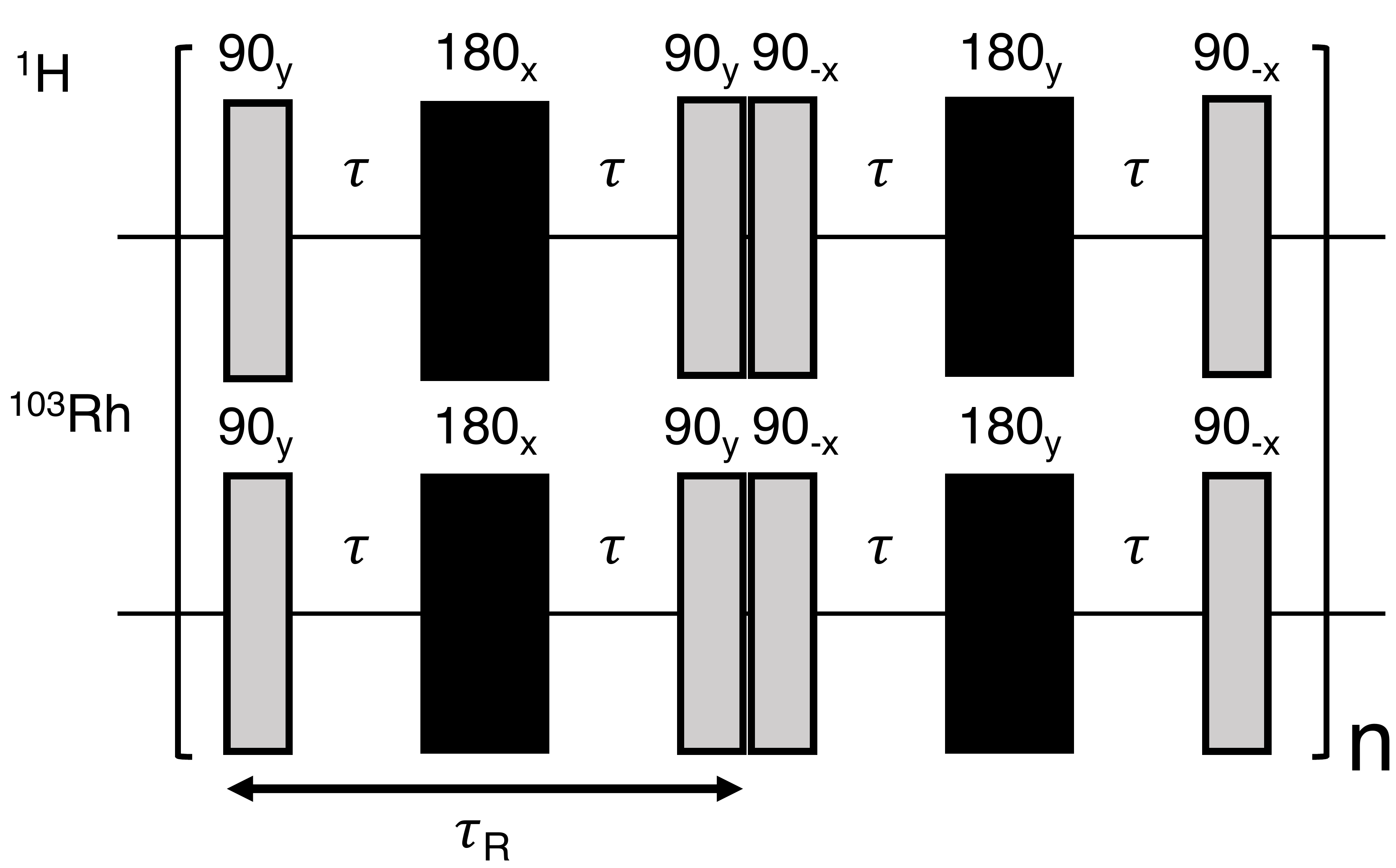}
\setlength{\belowcaptionskip}{-0pt}
\caption{\label{fig:DualPol} 
DualPol pulse sequence used for $\mathrm{^1H}$-$\mathrm{^{103}Rh}$ cross polarisation, and consisting of simultaneous PulsePol sequences~\cite{schwartz_robust_2018} on the two channels. Each PulsePol sequence is a repeating sequence of two R-elements. Each R-element has duration $\tau_R$, and is given by a composite 180\deg pulse~\cite{levitt_nmr_1979} with delays of duration $\tau$ between the pulses. The R-element duration should be short compared to the inverse of the relevant J-couplings. The black rectangles indicate BB1 composite $\pi$-pulses (equation~\ref{eq:composite180}).
}
\end{figure}
\subsubsection{DualPol Polarization Transfer Sequence}

The transfer of polarisation between \Rh and $^{1}$H  was achieved using the pulse sequence shown in figure~\ref{fig:DualPol}. This consists of repeating PulsePol sequences~\cite{schwartz_robust_2018,tratzmiller_pulsed_2021}, applied simultaneously to the $^{1}$H and \Rh radiofrequency channels. The PulsePol sequence consists of six phase-shifted radiofrequency pulses and four intervals $\tau$, and was originally developed for polarization transfer between electron and nuclear spins in the context of nitrogen-vacancy diamond magnetometry~\cite{schwartz_robust_2018}. It has also been shown to be effective for singlet-to-magnetization conversion~\cite{tratzmiller_pulsed_2021,sabba_symmetry-based_2022}, and has been interpreted in terms of symmetry-based recoupling theory~\cite{sabba_symmetry-based_2022}. For convenience, we refer to the "dual PulsePol" sequence in figure~\ref{fig:DualPol} as ``DualPol". 

 DualPol is an unusual example of a solution-state polarization transfer sequence combining (i) multiple-pulse averaging \cite{haeberlen_coherent_1968,zuiderweg_analysis_1990} and (ii) hard pulses separated by delays. The sequence provides robust polarization transfer even in the strong-coupling regime, 
 where the standard INEPT sequence breaks down~\cite{baishya_perfect_2014,williamson_one-_2000,koskela_aspects_2005,gil_accurate_2011,koskela_lr-cahsqc_2003}. That particular feature is not essential for the results described here. However, it is advantageous in other circumstances, as will be discussed in a future publication. 

 The repeating sequences of PulsePol and DualPol are composed of three-pulse elements of the form
 $90_y 180_x 90_x$, with the pulses separated by intervals $\tau$, and variants thereof. Each three-pulse sequence is therefore a ``windowed" version of a composite 180\deg pulse~\cite{levitt_nmr_1979}. We therefore call this three-pulse sequence a ``R-element", using notation originally introduced in the context of broadband heteronuclear decoupling~\cite{levitt_composite_1981}, and later adapted for symmetry-based recoupling sequences in solid-state NMR~\cite{carravetta_symmetry_2000}, and symmetry-based singlet-triplet conversion sequences in solution NMR~\cite{sabba_symmetry-based_2022}. In the  case of DualPol, there is no special constraint or matching condition on the duration $\tau_R$ of the R-element, except that it should be much shorter than the period of the relevant J-coupling, $\tau_R\ll\abs{^3J_\mathrm{RhH}}^{-1}$. Under these conditions, the average Hamiltonian~\cite{haeberlen_coherent_1968} generated by the DualPol sequence, for a heteronuclear 2-spin system, has the form
 \begin{equation}\label{eq:Hbar1}
     \overline{H}^{(1)} \simeq
     \kappa_\mathrm{DP} \times 2\pi J_{IS}\left(I_{x}S_{x} + I_{y}S_{y}\right)
 \end{equation}
 where the nuclides \Proton and \Rh are referred to as $I$ and $S$, respectively. The numbering convention for the average Hamiltonian terms starts with $1$ for the lowest-order approximation, in common with the symmetry-based recoupling literature~\cite{carravetta_symmetry_2000}. 
 The DualPol scaling factor is given, under suitable approximations, by $\kappa_\mathrm{DP}\simeq \frac{1}{2}$ in the limit of strong radiofrequency pulses. Equation~\ref{eq:Hbar1} corresponds to an anisotropic Hartmann-Hahn Hamiltonian \cite{glaser_homonuclear_1996}, indicating that the DualPol sequence exchanges $z$-magnetization components between the $I$-spins and $S$-spins. The theory and performance of the DualPol sequence will be discussed in more depth in a future paper.

In the experiments described here, all DualPol sequences used an R-element duration of $\tau_R = 5\mathrm{\,ms}$ and a repetition number of $n=10$. The total duration of each DualPol sequence was $T=2n\tau_R=100\mathrm{\,ms}$.

\begin{figure}[tbh]
\hspace*{-0.68cm}
\centering
\includegraphics[trim={4cm 0cm 8cm 0.8cm},clip,width=0.85\columnwidth]{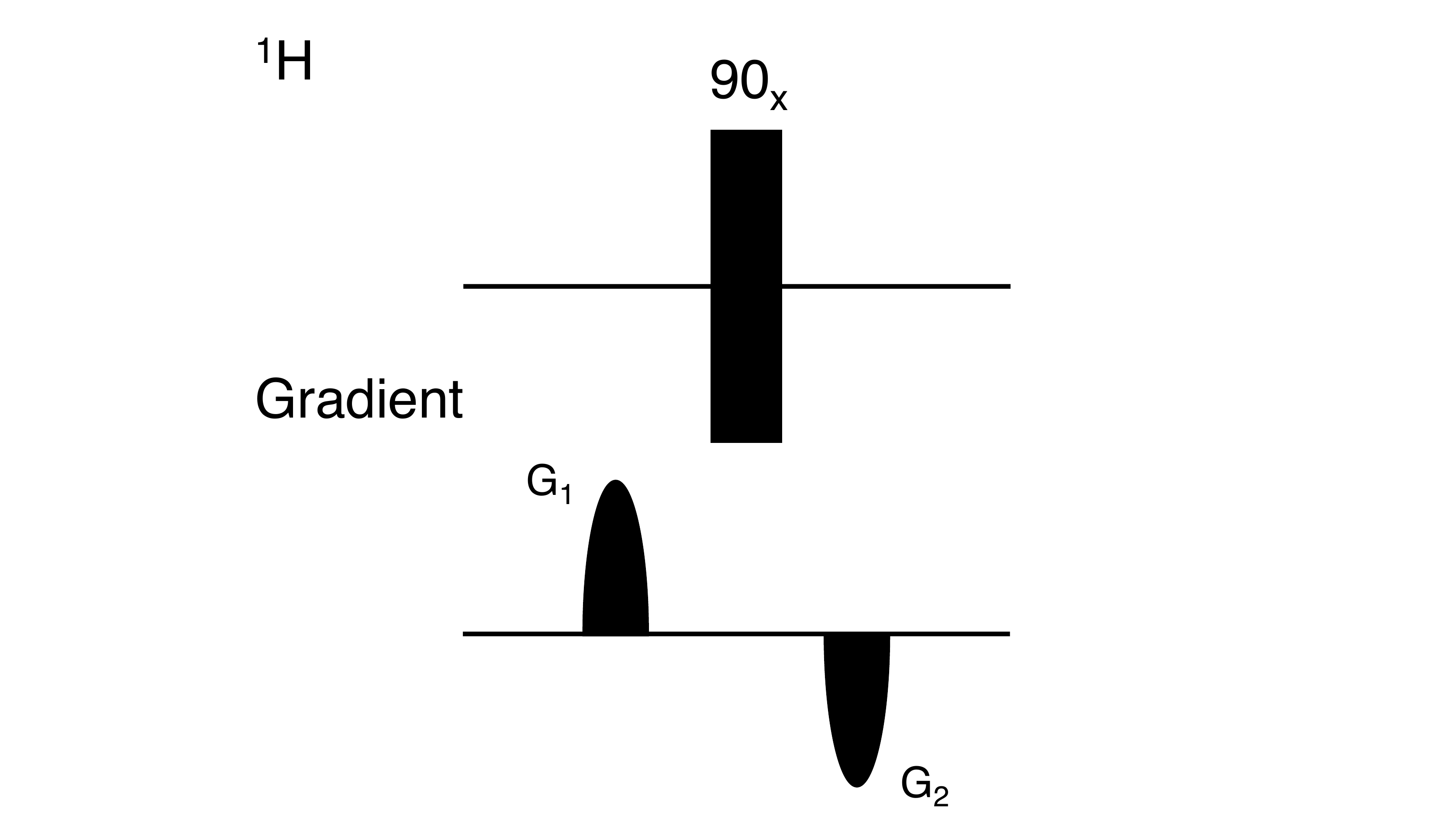}
\setlength{\belowcaptionskip}{-0pt}
\caption{\label{fig:ProtonDestructionFilter} 
Proton destruction filter for the removal of residual proton magnetisation. The gradient strengths are given by G$_1$=100$\%$ and G$_2$=-61.8$\%$ with respect to the maximum gradient strength 50 G cm$^{-1}$. Each gradient has a duration of 2~ms. The black rectangle indicates a BB1 composite $\pi/2$ pulse (equation~\ref{eq:composite90}).
}
\end{figure}
\subsubsection{
$^1$H Destruction Filter}

The \Proton destruction filter is shown in figure~\ref{fig:ProtonDestructionFilter}. The filter has the net effect of dephasing residual proton transverse and longitudinal magnetisation (which may be generated by accidental excitation, and recovery during the decay interval respectively).

\begin{figure}[tbh]
\hspace*{-0.68cm}
\centering
\includegraphics[trim={4cm 0cm 8cm 0.8cm},clip,width=0.85\columnwidth]{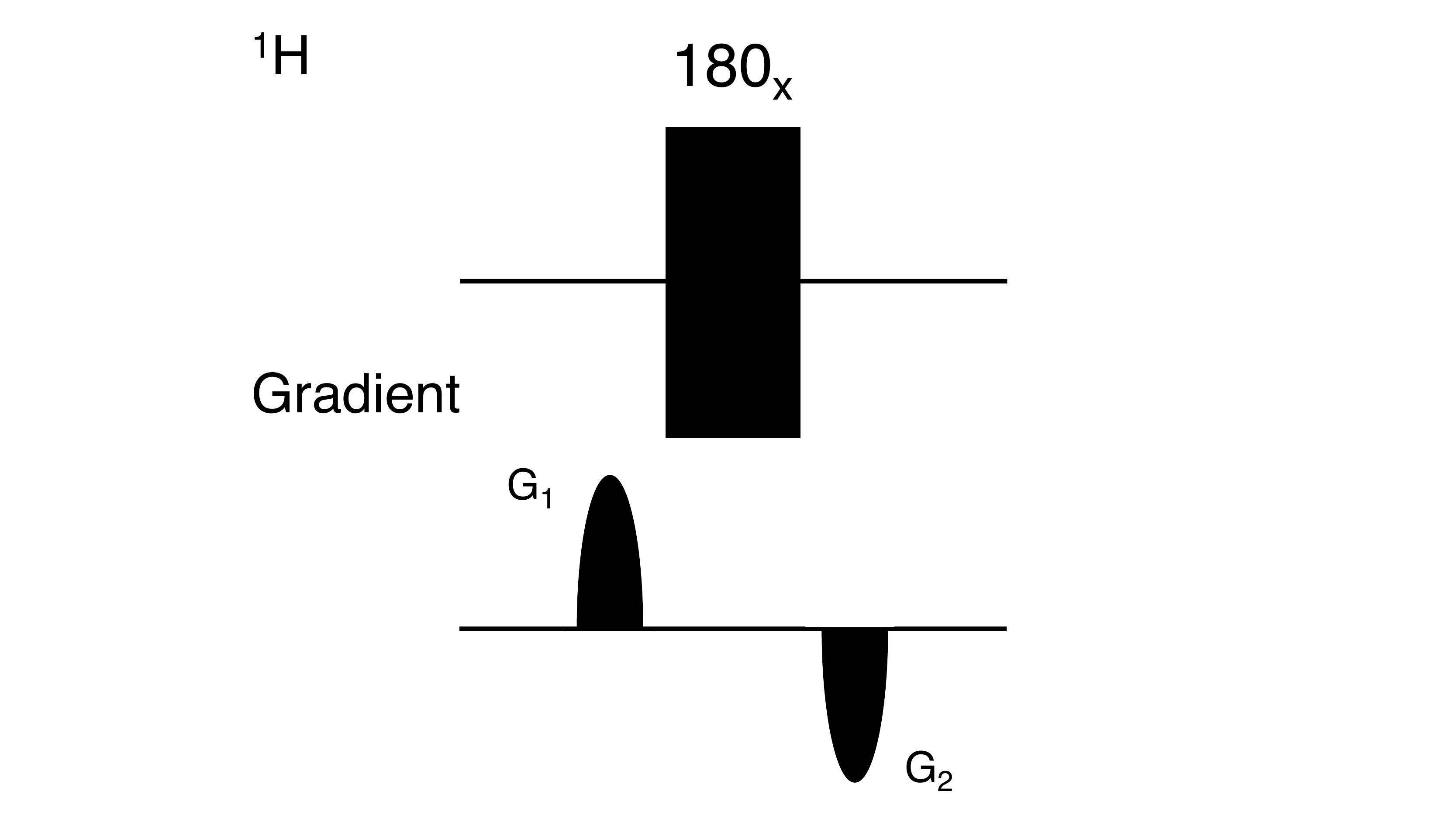}
\setlength{\belowcaptionskip}{-0pt}
\caption{\label{fig:ProtonZFilter} 
Proton z-filter for the selection of proton z-magnetisation, using bipolar gradients. The gradient strengths are given by G$_1$=$40\%$ and G$_2$=-40$\%$ with respect to the maximum gradient strength of 50~G cm$^{-1}$. Each gradient pulse has a duration of 2~ms. 
The black rectangle indicates a BB1 composite $\pi$-pulse (equation~\ref{eq:composite180}). }
\end{figure}
\subsubsection{$^1$H z-filter}
The z-filter for the selection of longitudinal \Proton magnetisation is shown in figure~\ref{fig:ProtonZFilter}. 
This employs a bipolar gradient scheme in order to reduce spectral distortions by eddy currents or residual gradient fields~\cite{wider_self-compensating_1994}.

\subsection{Solid-state NMR}

Solid state CPMAS $^{13}$C NMR was performed using a 4~mm Bruker probe at 14.1~T and $\sim$303~K.

\subsection{Computational Chemistry}

Quantum chemical geometry optimisation and shielding tensor calculations for the rhodium formate complex axially ligated by solvent THF molecules were performed using the ORCA program package version 5.0.3~\cite{neese_orca_2020}. \Rh shielding tensors were computed at the TPSSh/SARC-ZORA-TZVPP level of theory.

\begin{figure}[tbh]
\hspace*{-0.68cm}
\centering
\includegraphics[trim={1.5cm 0cm 1.5cm 0cm},clip,width=1\columnwidth]{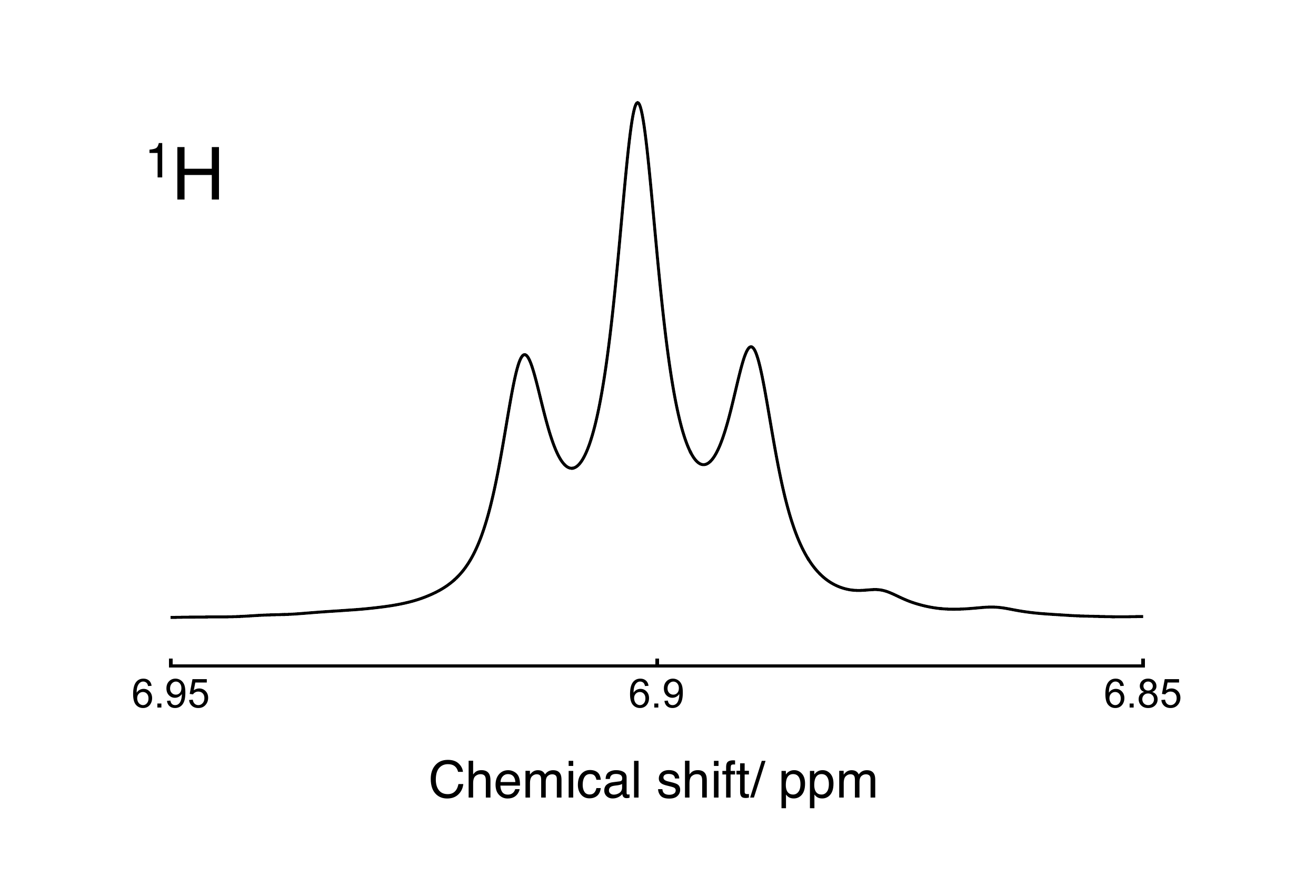}
\setlength{\belowcaptionskip}{-0pt}
\caption{\label{fig:ProtonSpectrum} 
\Proton spectrum of a \hl{$\sim$10 mM solution} of rhodium formate in THF-d$_8$, acquired at 9.4 T and at 298K \hl{in a single scan. Exponential line broadening (0.75 Hz) was applied.} 
}
\end{figure}
\section{Results}
\subsection{NMR Spectra}
\subsubsection{Solution-state $^1$H Spectrum} 
The rhodium formate \Proton spectrum features a single formate \Proton resonance split into a 1:2:1 triplet by coupling to the pair of magnetically equivalent \Rh nuclei (figure~\ref{fig:ProtonSpectrum}). The three-bond \Proton-\Rh J-coupling is estimated to be $|^3J_\mathrm{RhH}|=4.7\pm0.1$~Hz.

\begin{figure}[b]
\hspace*{-0.68cm}
\centering
\includegraphics[trim={0cm 0cm 0cm 0 cm},clip,width=1\columnwidth]{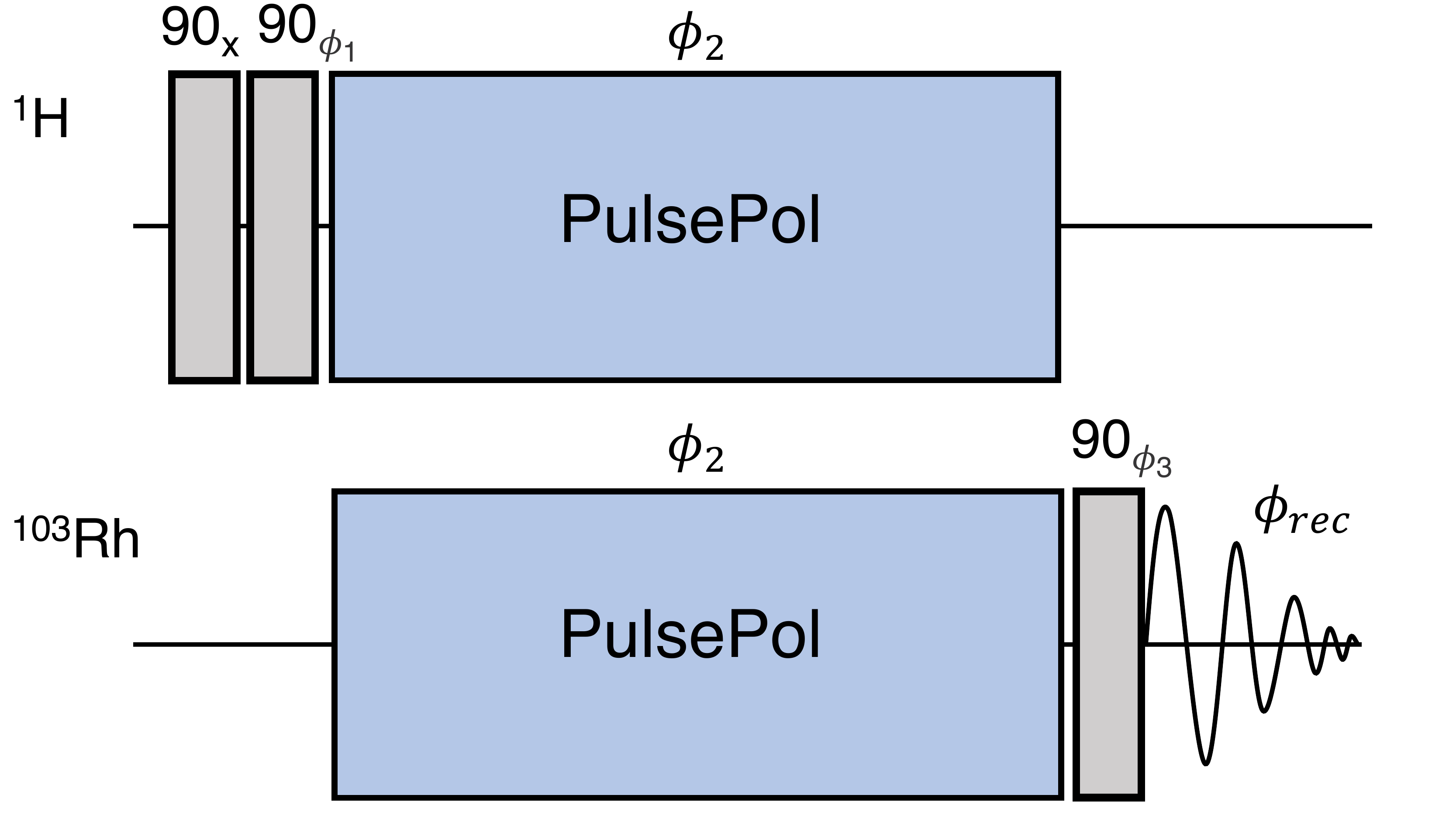}
\setlength{\belowcaptionskip}{-0pt}
\caption{\label{fig:ProtonEnhancedRhodiumSpectra} 
Pulse sequence for the acquisition of $\mathrm{^1H}$ enhanced $\mathrm{^{103}Rh}$ spectra. A 16-step phase cycle is used where $\phi_1 = [{-x,x,-x,x}]$, $\phi_2 = [{x,x,-x,-x}]$, $\phi_3 = [x,x,x,x,y,y,y,y,-x,-x,-x,-x,-y,-y,-y,-y]$ and the receiver $\phi_{rec}$ = [x,-x,x,-x,y,-y,y,-y,-x,x,-x,x,-y,y,-y,y] all of which combine to suppress ringing artefacts on the \Rh channel.
}
\end{figure}
\begin{figure}[tbh]
\hspace*{-0.68cm}
\centering
\includegraphics[trim={0cm 0.3cm 0.8cm 8.5cm},clip,width=1\columnwidth]{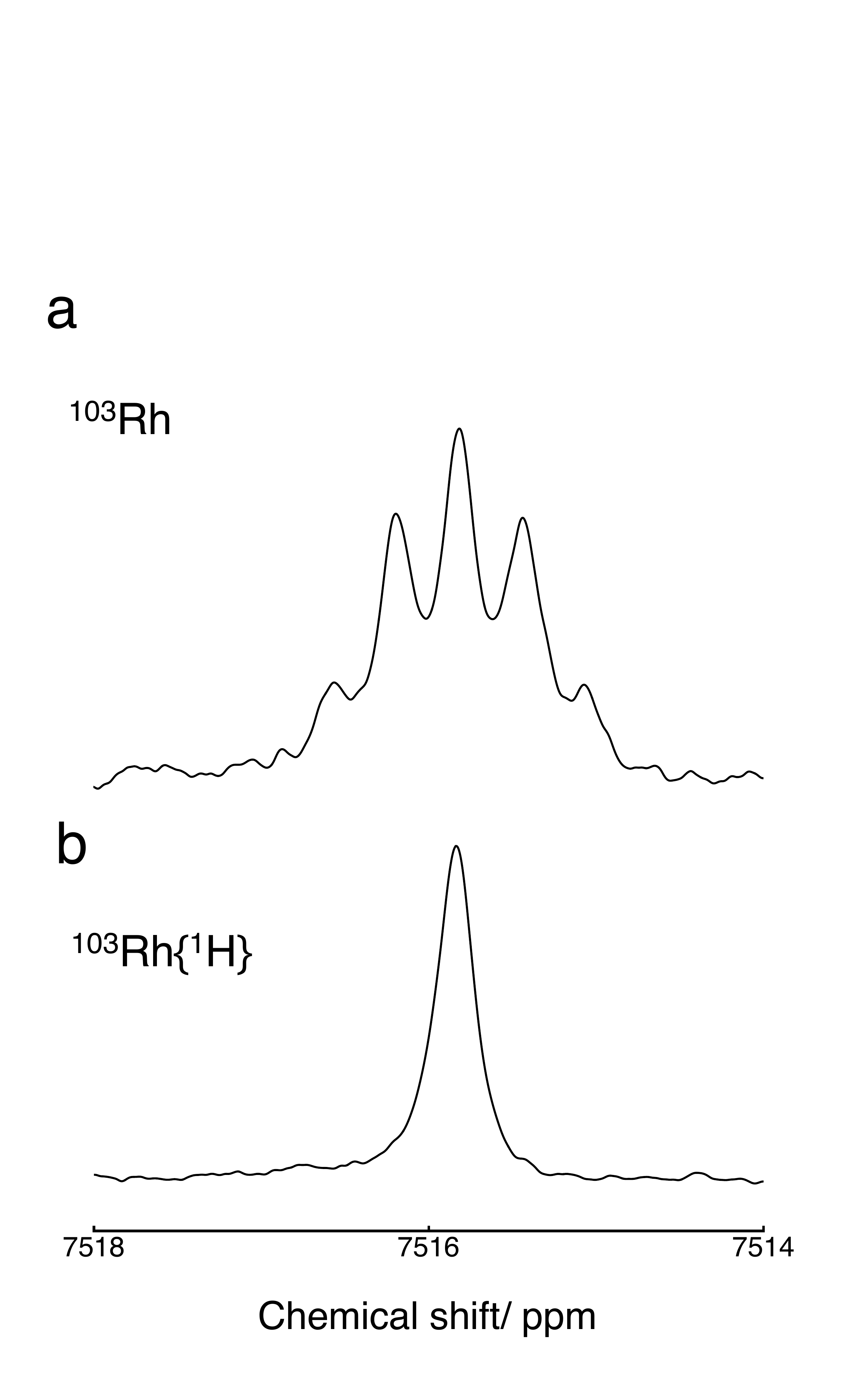}
\setlength{\belowcaptionskip}{-0pt}
\caption{\label{fig:CombinedRhodiumSpectra} 
(a) \Rh spectrum of a \hl{$\sim$10 mM solution} of rhodium formate in THF-d$_8$ scaled 2.5 times, acquired using 128 scans at 9.4~T and at 298~K using the pulse sequence in figure~\ref{fig:ProtonEnhancedRhodiumSpectra}. 
(b) \Proton-decoupled \Rh spectrum acquired using 128 scans at 9.4~T and at 298~K using the pulse sequence in figure~\ref{fig:ProtonEnhancedRhodiumSpectra} with continuous-wave \Proton decoupling during signal acquisition. 
\hl{Acquisition time for each spectrum was ~1 hour. Exponential line broadening (1 Hz) was applied to each spectrum.}}
\end{figure}
\subsubsection{Solution-state $^{103}$Rh Spectra} 

The sequence shown in figure~\ref{fig:ProtonEnhancedRhodiumSpectra} was used for the acquisition of directly-detected \Rh spectra, enhanced by polarization transfer from \Proton nuclei. 
After an initial pair of 90\deg pulses, used for the suppression of ringing artefacts (see below), the DualPol sequence transfers z-magnetization from the \Proton to the \Rh nuclei, exploiting the form of the DualPol average Hamiltonian (equation~\ref{eq:Hbar1}).  
The resultant \Rh z-magnetization is converted into observable transverse magnetization by a final 90\deg pulse. The \Rh NMR signal is enhanced by a factor of up to $\abs{\gamma_I/\gamma_S}\sim31$, relative to that induced by a single 90\deg pulse applied to \Rh nuclei in thermal equilibrium. 

Ringing artifacts are strongly suppressed by a phase-cycled pair of 90\deg pulses on the proton channel, before the polarization transfer takes place. The signs of the \Proton magnetization and the \Rh receiver are simultaneously inverted in successive scans. Since the phases of the ringing are correlated with the phases of the pulses on the \Rh channel, the ringing is strongly suppressed in the  \Rh spectrum. Further suppression of ringing is achieved by additional phase cycling of the PulsePol blocks. The sign of the \Rh magnetization is invariant under global phase shifts of the DualPol sequence, while the ringing contribution is phase-correlated and largely cancels out. Similar logic has been used to design excitation schemes for ringing suppression in homonuclear NMR experiments \cite{zhang_elimination_1990,wang_triple-pulse_2021}.

The rhodium formate \Rh spectrum features a single \Rh resonance split into a 1:4:6:4:1 pentet by couplings to the four equivalent \Proton nuclei on the formate ligands (figure~\ref{fig:CombinedRhodiumSpectra}(a). The three-bond \Proton-\Rh J-coupling is estimated to be $|^3J_\mathrm{RhH}|=4.7\pm0.1\mathrm{\,Hz}$, in agreement with the \Proton spectrum.
The \Rh resonances collapse into a single peak centred at 7516~ppm upon \Proton decoupling (figure~\ref{fig:CombinedRhodiumSpectra}(b)). 

\hl{The $^{103}$Rh resonances are broadened by the short $^{103}$Rh $T_2$ (see figure}-\ref{fig:RhT2}).  

\begin{figure}[tbh]
\hspace*{-0.68cm}
\centering
\includegraphics[trim={5.75cm 1cm 7.75cm 3cm},clip,width=1\columnwidth]{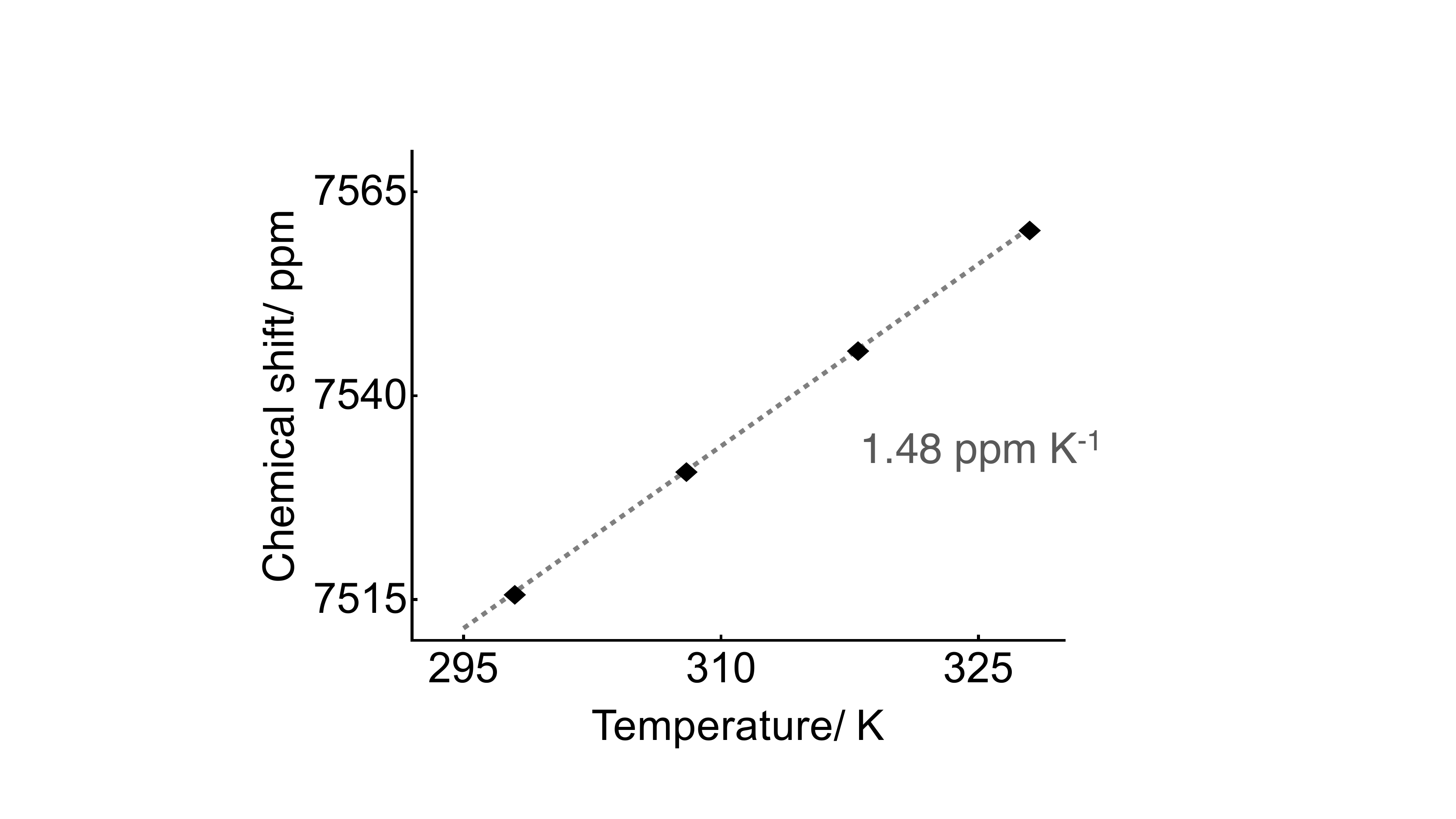}
\setlength{\belowcaptionskip}{-0pt}
\caption{\label{fig:RhvsT} 
\Rh
chemical shift of rhodium formate dissolved in THF-d$_8$ at 9.4~T, as a function of temperature. The chemical shifts are referenced to $\Xi\left(^{103}\rm{Rh}\right)$ = 3.16$\%$.
}
\end{figure}

The \Rh chemical shift is temperature-dependent (see Figure~\ref{fig:RhvsT}). The
temperature-dependence of the \Rh chemical shift is approximately linear over the relevant temperature range, with a gradient of $\sim1.48\mathrm{\ ppm\,K^{-1}}$. This is in general agreement with observations on similar Rh complexes~\cite{carlton_chapter_2008,calo_triple_2021}.

\subsubsection{Solid-state \Cth NMR}\label{sec:CthNMR}

\begin{figure}[tbh!]
\hspace*{-0.68cm}
\centering
\includegraphics[width=\columnwidth]{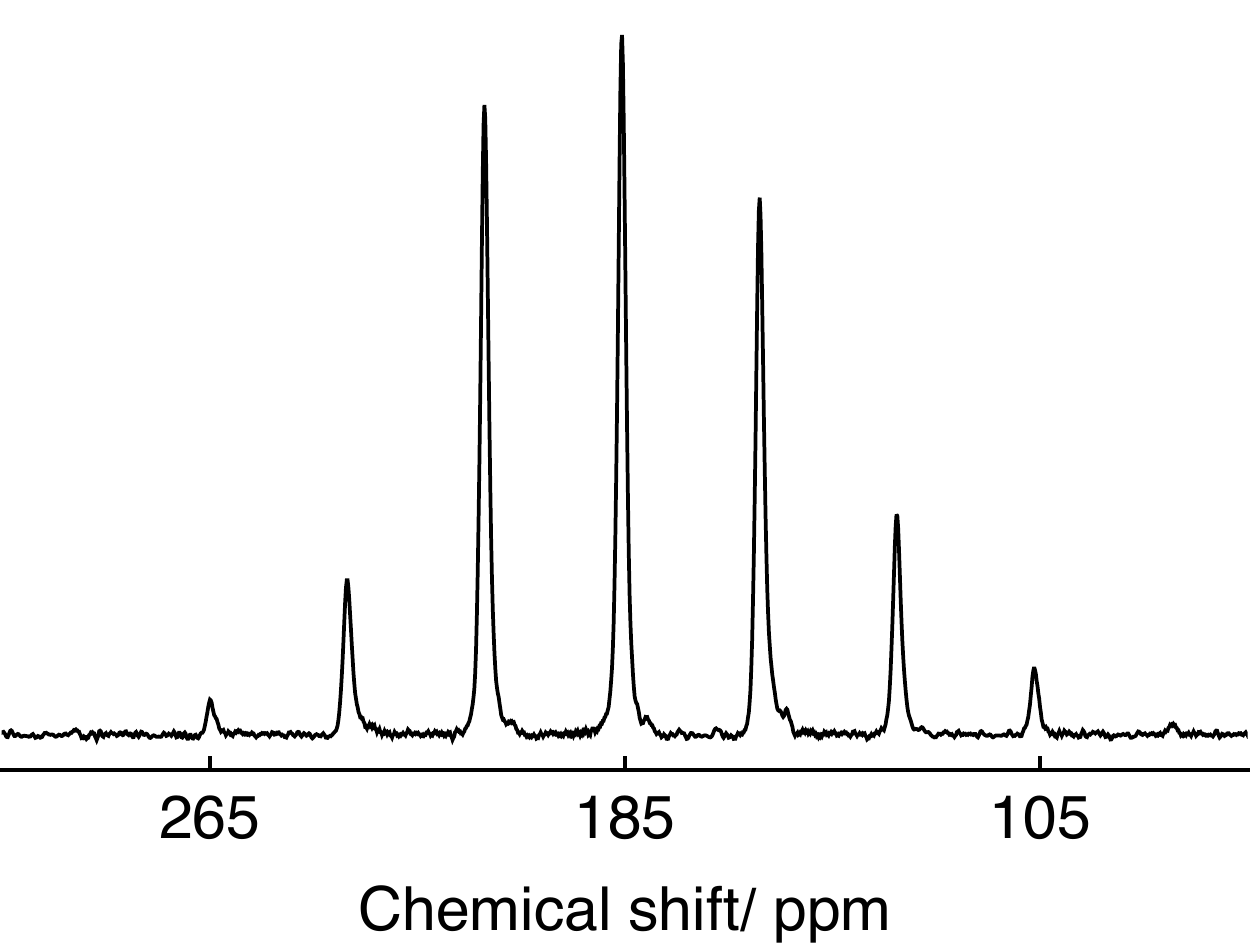}
\setlength{\belowcaptionskip}{-0pt}
\caption{\label{fig:13CssNMR} 
Rhodium formate $^{13}$C$\{^1\mathrm{H}\}$ solid-state CPMAS\cite{meier_cross_1992} NMR spectrum obtained at a spinning frequency of 4~kHz acquired using 2048 scans at 14.1 T and at ~303K. \hl{The chemical shift was referenced to admantane. The contact time was 160~$\mu$s. The recycle delay was 3 s. $\sim$150 mg of sample was used. Further details of the pulse sequence are provided in the Supporting Information.}
}
\end{figure}

The chemical shift anisotropy (CSA) of the formate \Cth nuclei was estimated by magic-angle-spinning NMR experiments on rhodium formate solid (figure~\ref{fig:13CssNMR}). 

\hl{The estimated eigenvalues of the traceless, symmetric (rank-2) part of the shielding tensor are as follows: $\sigma^{(2)}_{xx}=65.1 \mathrm{\,ppm}$, $\sigma^{(2)}_{yy}=5.5\mathrm{\,ppm}$, and $\sigma^{(2)}_{zz}=-70.7 \mathrm{\,ppm}$. This corresponds to the following Frobenius norm of the rank-2 $^{13}$C shielding tensor:}

\begin{align}
\label{eq:normsigma}
||\boldsymbol{\sigma}^{(2)}||
(^{13}\mathrm{C})
&= \{(\sigma^{(2)}_{xx})^2+(\sigma^{(2)}_{yy})^2+(\sigma^{(2)}_{zz})^2\}^{1/2}
\nonumber\\&=
 96.3 \pm 1.0\mathrm{\ ppm}
\end{align}

\begin{figure}[tbh]
\hspace*{-0.68cm}
\centering
\includegraphics[trim={0cm 0cm 0cm 0cm},clip,width=1\columnwidth]{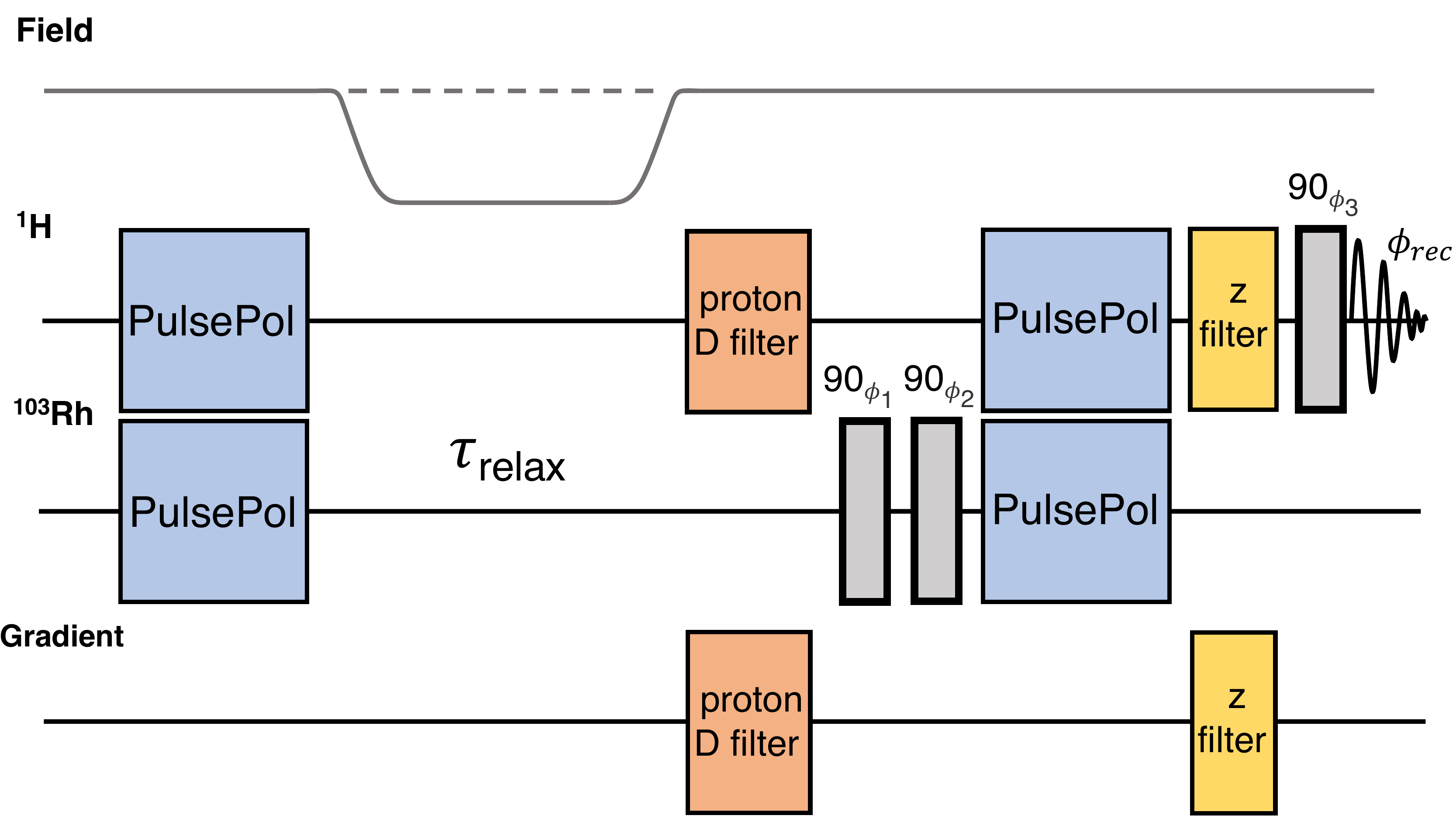}
\setlength{\belowcaptionskip}{-0pt}
\caption{\label{fig:IndirectT1sequence} 
Sequence used for the indirect measurement of rhodium T$_1$ through \Proton NMR signals. Phase cycles are given by $\phi_1 = [{x,x,-x,-x}]$, $\phi_2 = [{-x,x,-x,x}]$, $\phi_3=[x,x,x,x,y,y,y,y,-x,-x,-x,-x,-y,-y,-y,-y]$ and the receiver $\phi_{rec}=[x,-x,-x,x,y,-y,-y,y,-x,x,x,-x,-y,y,y,-y]$. 
The optional shuttling of the sample to low field, and back again, during the interval $\tau_\mathrm{relax}$, is indicated.
}
\end{figure}
\subsection{Relaxation Times}
\subsubsection{\Proton-Detected \Rh $T_1$}

\Rh $T_1$ relaxation time constants were measured indirectly through \Proton NMR signals using the sequence shown in figure~\ref{fig:IndirectT1sequence}. DualPol is used to transfer z-magnetization from the \Proton nuclei to the \Rh nuclei, and allowed to relax towards equilibrium during the relaxation interval $\taurelax$. For field-dependent relaxation measurements, the sample is shuttled to a region of lower magnetic field during this interval, and back again. 
A proton destruction filter is applied to eliminate any residual proton magnetisation, \hl{such as that generated during $\taurelax$ through longitudinal relaxation towards equilibrium}. Remaining \Rh z-magnetisation, selected for by the two 90\deg pulses, is now transferred back to \Proton z-magnetisation by a second DualPol block and is selected for by a proton z-filter. A final \Proton 90\deg pulse generates observable \Proton transverse magnetization. The sequence is repeated with variation of $\taurelax$ in order to follow the equilibration of longitudinal \Rh magnetization.

\begin{figure}[tbp]
\hspace*{-0.68cm}
\centering
\includegraphics[trim=1.2cm 0cm 4.5cm 0.5cm,clip,width=0.95\columnwidth]{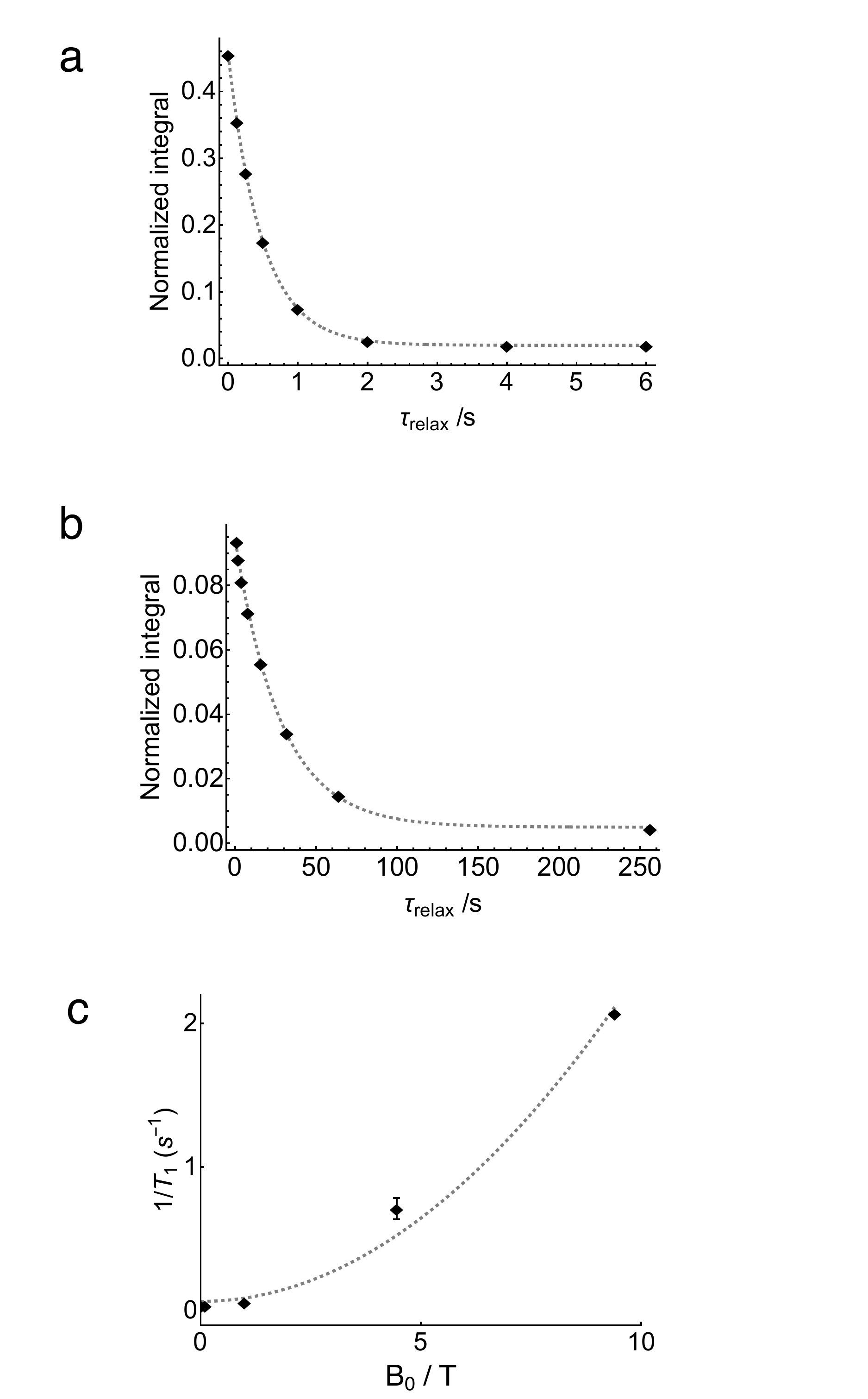}
\setlength{\belowcaptionskip}{-0pt}
\caption{\label{fig:RhT1} 
(a) Decay curve for \Rh longitudinal magnetization at a field of 9.4~T, obtained using the pulse sequence in figure~\ref{fig:IndirectT1sequence}, but without shuttling the sample to low field. \hl{The data was acquired in $\sim$20 minutes.} The integrals are normalised against the \Proton spectrum obtained by a single \Proton 90\deg pulse applied to a system in thermal equilibrium at 9.4 T. The data fits well to an exponential decay with time constant $T_1=0.483\pm0.002\mathrm{\,s}$. 
(b) Decay curve for \Rh longitudinal magnetization at a field of 1~mT, obtained using the pulse sequence in figure~\ref{fig:IndirectT1sequence}, including the shuttling of the sample to low field. The data fits well to an exponential decay with time constant $T_1=28.2\pm1.2\mathrm{\,s}$. 
(c) $^{103}$Rh relaxation rate constant $T_1^{-1}$ as a function of magnetic field strength. The dashed line shows the quadratic function $T_1^{-1}(B)=T_1^{-1}(0)+a B^{2}$, where $T_1^{-1}(0) = 0.065\pm 0.038\mathrm{\,s^{-1}}$ and $a=0.023\pm 0.001\mathrm{\,s^{-1}\,T^{-2}}$. 
}
\end{figure}

The trajectory of indirectly-detected \Rh z-magnetization in a field of $9.4~\mathrm{T}$ is shown in figure~\ref{fig:RhT1}(a). 
The trajectory fits well to a single-exponential decay with time constant $T_1(^{103}\mathrm{Rh})=0.483\pm0.002\mathrm{\ s}$. A trajectory in the low magnetic field of $\mathrm{1\ mT}$ is shown in figure~\ref{fig:RhT1}(b). This was produced by shuttling the sample to low magnetic field during the interval $\taurelax$. 
The relaxation process is much slower in low field, with a time constant of $T_1(^{103}\mathrm{Rh})=28.2\pm1.2\mathrm{\ s}$.

The rhodium $T_1^{-1}$ increases approximately quadratically with the magnetic field strength $B$, as shown in figure~\ref{fig:RhT1}(c). The field-dependent relaxation rate constant is a reasonable fit to the quadratic function 
$T_1^{-1}(B)=T_1^{-1}(0)+aB^{2}$, where $T_1^{-1}(0) = 0.065\pm 0.04\mathrm{\,s^{-1}}$ and $a=0.023\pm 0.001\mathrm{\,s^{-1}\,T^{-2}}$.

\begin{figure}[tbh]
\hspace*{-0.68cm}
\centering
\includegraphics[trim={0cm 0cm 0cm 0cm},clip,width=1\columnwidth]{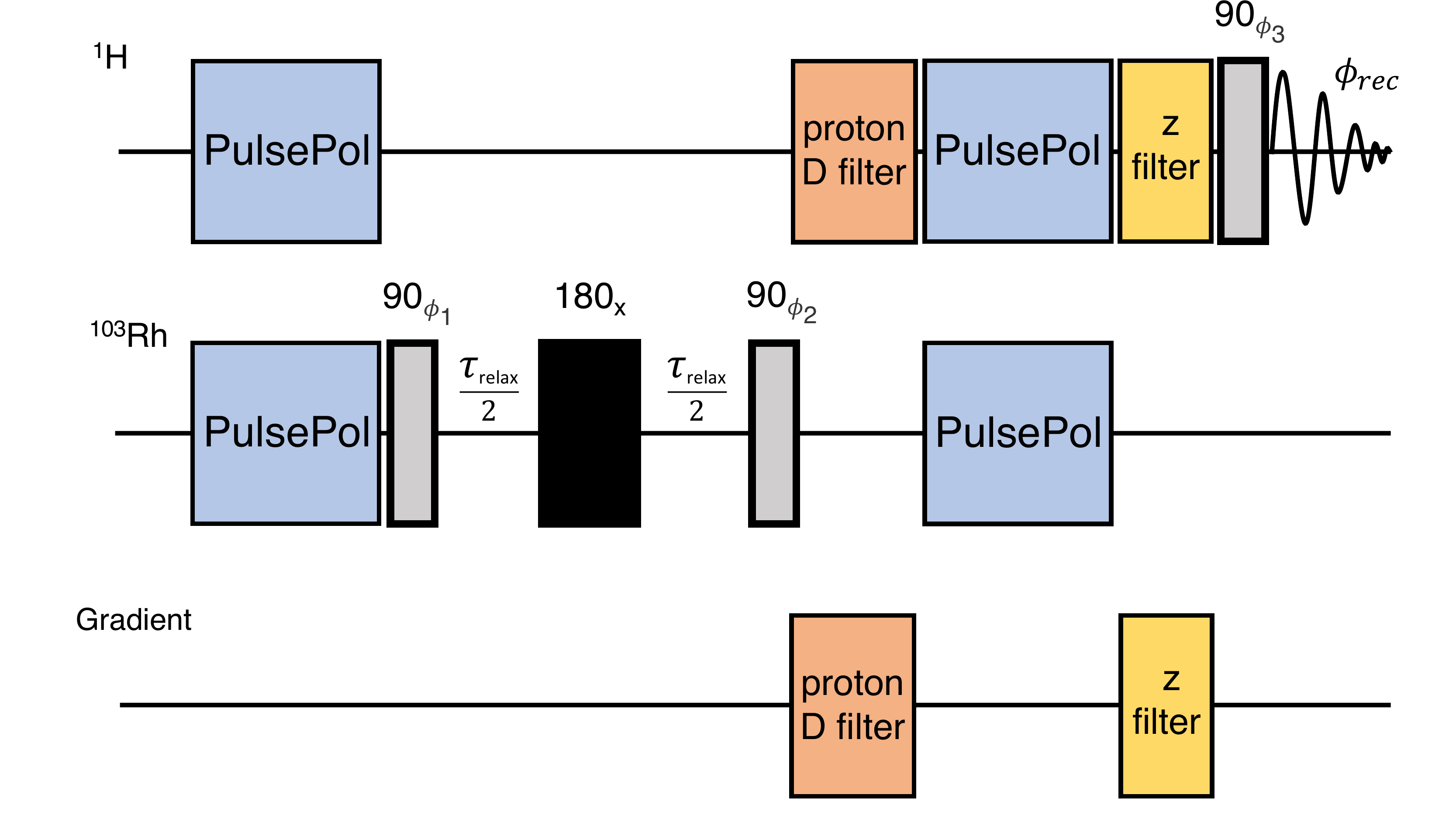}
\setlength{\belowcaptionskip}{-0pt}
\caption{\label{fig:IndirectT2sequence} 
Sequence used for the indirect measurement of rhodium T$_2$ with detection on protons. Phase cycles are given by $\phi_1 = [{x,x,-x,-x}]$, $\phi_2 = [{-x,x,-x,x}]$, $\phi_3=[x,x,x,x,y,y,y,y,-x,-x,-x,-x,-y,-y,-y,-y]$ and the receiver $\phi_{rec}$=[x,-x,-x,x,y,-y,-y,y,-x,x,x,-x,-y,y,y,-y]. The black rectangle indicates a BB1 composite $\pi$-pulse (equation~\ref{eq:composite180}). 
}
\end{figure}
\begin{figure}[b]
\hspace*{-0.68cm}
\centering
\includegraphics[trim={4cm 0cm 5cm 1cm},clip,width=1\columnwidth]{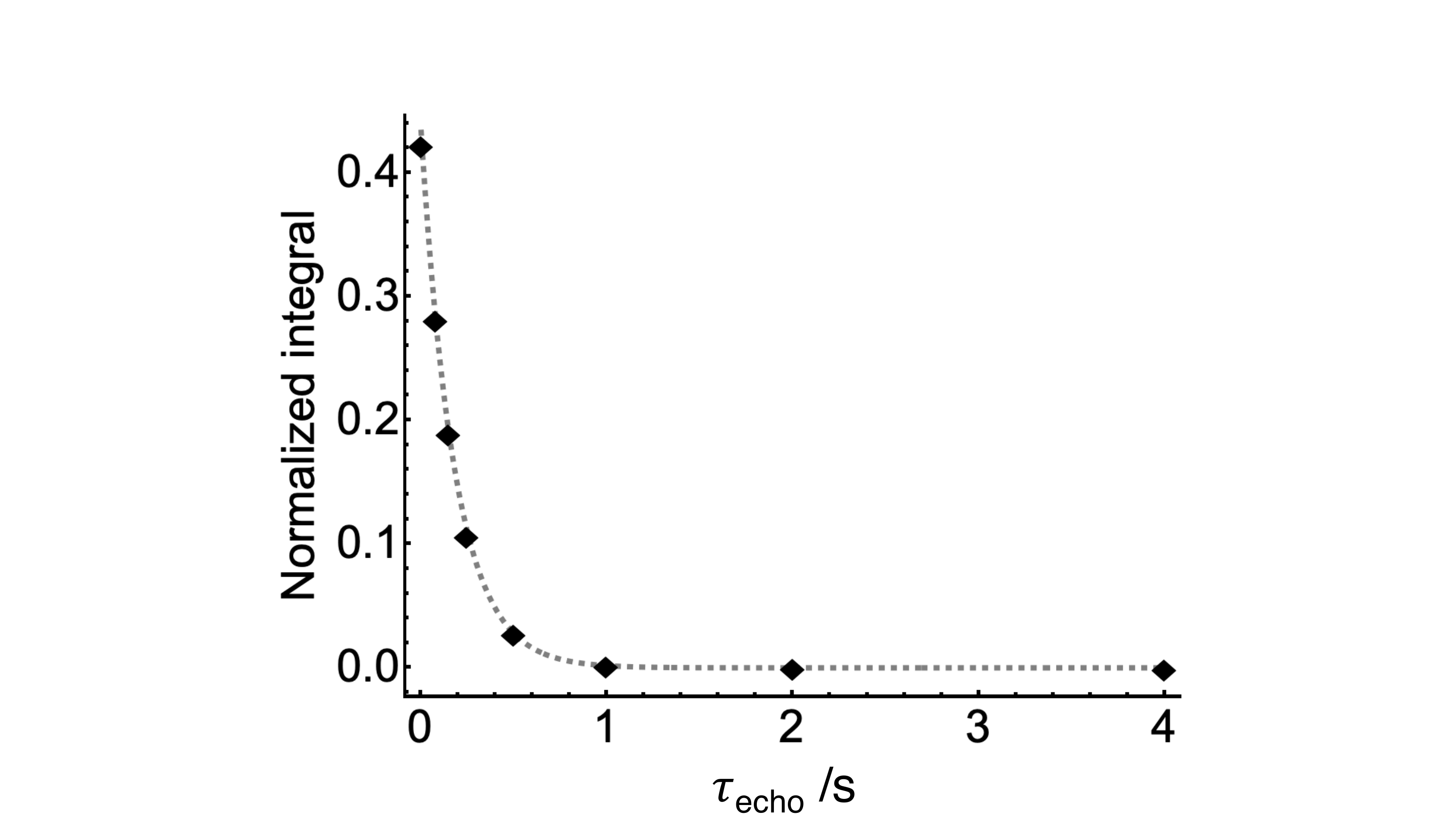}
\setlength{\belowcaptionskip}{-0pt}
\caption{\label{fig:RhT2} 
Decay curve for $^{103}$Rh transverse magnetization at a field of 9.4~T, obtained using the pulse sequence in figure~\ref{fig:IndirectT2sequence}. 
The data fits well to an exponential decay with time constant $T_2=0.181\pm0.001\mathrm{\,s}$. The integrals are normalised against the \Proton spectrum obtained by a single \Proton 90\deg pulse applied to a system in thermal equilibrium at 9.4 T.
}
\end{figure}

\subsubsection{\Proton-Detected \Rh $T_2$}

The sequence shown in figure~\ref{fig:IndirectT2sequence} was used to measure the \Rh spin-spin relaxation time constant $T_2$ in high magnetic field. 

Conversion of \Proton z-polarization to \Rh z-polarization is achieved via DualPol. \Rh transverse magnetisation is generated by a 90\deg pulse and allowed to decay during the subsequent spin echo of duration $\tau_\mathrm{echo}$. The ensuing 90\deg \Rh pulse returns the remaining transverse \Rh magnetisation back to longitudinal \Rh polarisation. A \Proton destruction filter destroys any residual \Proton magnetisation before another DualPol cross-polarisation block transfers \Rh z-magnetisation back to \Proton z-magnetization. The \Proton z-filter selects for \Proton z-magnetization before the \Proton signal is induced by the final 90\deg \Proton pulse. The pulse sequence is repeated varying the echo delay $\tau_\mathrm{echo}$ in order to follow the decay of \Rh transverse magnetization. 

The trajectory of indirectly-detected \Rh transverse magnetization in a field of $9.4~\mathrm{T}$ is shown in figure~\ref{fig:RhT2}. The trajectory fits well to a single-exponential decay with time constant $T_2($\Rh$\!)=0.181\pm0.001\mathrm{\ s}$. Note that the measured value of $T_2$ is much smaller than $T_1$ under the same conditions. 

\subsubsection{\Cth inversion-recovery}
\hl{
As discussed below, the rotational correlation time $\tau_c$ of the rhodium formate complex may be estimated by a study of the $^{13}$C longitudinal relaxation. This data was obtained by an indirect detection method exploiting the scalar-coupled formate protons, as described in the Supporting Information. The inversion-recovery data fits well to a single-exponential recovery with a time constant of $2.64\pm0.13\mathrm{\,s}$ for a solution in THF-d$_8$, in a magnetic field of $9.4$~T. However, as described below, the inversion-recovery curve for the $^{13}$C magnetization is best analyzed using a bi-exponential relaxation model.
}

\section{\label{sec:Discussion}Discussion}

As shown in figure~\ref{fig:RhT1}(c), the \Rh relaxation rate constant $T_1^{-1}$ has a quadratic dependence on magnetic field $B$, with an additional zero-field contribution of $T_1^{-1}(0)=0.0653\pm 0.0383\mathrm{\,s}^{-1}$. The quadratic field dependence is consistent with a dominant chemical shift anisotropy (CSA) relaxation mechanism, as is commonly observed for the \Rh NMR of rhodium complexes~\cite{cocivera_nmr_1982,carlton_chapter_2008}.

It is difficult to estimate the \Rh chemical shift anisotropy by solid-state NMR. The small magnetogyric ratio of \Rh and the very large CSA value make solid-state \Rh NMR very difficult. Our attempts to use the PROSPR method~\cite{jaroszewicz_sensitivity_2021} to observe the \Rh spectrum indirectly in the solid state, by saturation transfer to the \Proton nuclei, were also unsuccessful. 
\hl{This is likely due to the very small dipole-dipole couplings }\hl{between $^1$H and $^{103}$Rh nuclei in this complex, which greatly inhibits dipolar-mediated polarization transfer in the solid state.}

The symmetry of the complex indicates that the \Rh CSA tensors should have uniaxial symmetry ($\eta=0$) with their unique principal axis along the Rh-Rh bond. This property is assumed in the following discussion. 

Although the \Rh CSA may not be measured directly, it is possible to estimate it by a combination of field-dependent \Rh and \Cth $T_1$ measurements. The compact cage structure of the rhodium formate complex (figure~\ref{fig:Rhformatestruc}) suggests that, to a good approximation, the complex tumbles in solution as a near-rigid body, with a common rotational correlation time $\tau_c$ for all spin interactions. This approximation allows a correlation time estimate from \Cth NMR to be applied in the context of \Rh NMR. 

A \Cth nucleus of rhodium formate experiences two strong anisotropic interactions: the \Cth-\Proton dipole-dipole coupling with the directly-bonded hydrogen nucleus, and the \Cth chemical shift anisotropy. \hl{For point nuclei (i.e. ignoring the spatial spread of the nuclear wavefunctions), the 
} \Cth-\Proton dipole-dipole coupling constant is given by $b_{CH}=-(\mu_0/4\pi)\hbar \gamma_{C}\gamma_{H} r_{CH}^{-3}$, where $r_{CH}$ is the \Cth-\Proton internuclear distance~\cite{levitt_spin_2007}.
Quantum chemical calculations\cite{neese_orca_2020} (see SI) predict an internuclear \Cth-\Proton distance of 1.097 Å, corresponding to a dipole-dipole coupling constant of $b_{CH}=-2\pi\times 22.8$~kHz. \hl{However, solid-state NMR studies have shown that the true dipole-dipole coupling is weakened by the angular spread of the $^1$H wavefunctions, associated with the zero-point librational motion of the C-H bonds}~\cite{nakai_influence_1989}. \hl{In the calculations below, we therefore assume a $^{13}$C-$^1$H dipole-dipole coupling constant of $b_{CH}=-2\pi\times (20.4 \pm 0.5)$~kHz.}

\hl{For isolated $^{13}$C-$^1${H} spin systems in the extreme narrowing approximation (fast tumbling), the theoretical recovery of $^{13}$C longitudinal magnetization $M_z(t)$ after perturbation from equilibrium at time $t=0$ is expected to follow the biexponential curve}
\begin{align}
\label{eq:Mzrec}
    M_z(t) =
M_z^\mathrm{eq}+
&\left(
    M_z(0) - M_z^\mathrm{eq}
\right)\times
\nonumber\\&
\frac{1}{2}\!\left(
    \exp\{
-(\tfrac{1}{2}b_{CH}^2+
\tfrac{1}{5}\omega_\mathrm{CSA}^2)   \tau_c
    \}
\right.
\nonumber\\
&
+\left.
    \exp\{
-(\tfrac{3}{2}b_{CH}^2+
\tfrac{1}{5}
\omega_\mathrm{CSA}^2)   \tau_c
    \}
    \right)
\end{align}
\hl{where $M_z^\mathrm{eq}$ is the thermal equilibrium $^{13}$C magnetization, and $\omega_\mathrm{CSA}$ is defined as follows:}
\begin{equation}
\omega_\mathrm{CSA} = 
  -\gamma_C B^0 
  ||\boldsymbol{\sigma}^{(2)}||
\end{equation}
\hl{where $||\boldsymbol{\sigma}^{(2)}||$ is the norm of the $^{13}$C shielding tensor, as defined in equation}~\ref{eq:normsigma}.
\hl{The biexponential form of equation}~\ref{eq:Mzrec} \hl{is due to $^1$H-${13}$C cross-relaxation during the magnetization recovery}~\cite{Solomon_NOE_1955, alger_carbon13_1972,kowalewski_book_2018}.

\hl{In a magnetic field of 9.4~T, the $^{13}$C CSA, as estimated by $^{13}$C solid-state NMR (section}~\ref{sec:CthNMR})\hl{, corresponds to an interaction strength of $\omega_\mathrm{CSA} \simeq 2\pi\times(9.7\pm 0.1)\mathrm{\,kHz}$. By fitting the experimental $^{13}$C inversion-recovery trajectory to an equation of the form in eq.}~\ref{eq:Mzrec}\hl{, we obtain the following estimate of the rotational correlation time for the rhodium formate paddlewheel complex in THF-d$_8$ solution at $298\mathrm{\ K}$: $\tau_c\simeq 24.5\pm 1.5\mathrm{\,ps}$. 
}
The \Rh relaxation may now be analyzed using the estimate of $\tau_c$ from the \Cth data. As shown in figure~\ref{fig:RhT1}, the \Rh $T_1^{-1}$ relaxation rate constant is well-described by the function $T_1^{-1}(B)=T_1^{-1}(0)+a B^{2}$, with the field-independent term $T_1^{-1}(0) = 0.065\pm 0.038\mathrm{\,s^{-1}}$, and the quadratic coefficient $a=0.023\pm 0.001\mathrm{\,s^{-1}\,T^{-2}}$. 

The quadratic field-dependent term may be ascribed to the CSA mechanism. In the extreme narrowing approximation (fast tumbling), the CSA contribution to the $T_1^{-1}$ relaxation rate constant for \Rh is given by~\cite{kowalewski_book_2018}
\begin{equation}\label{eq:RhT1theory}
\left(T_{1}(^{103}\mathrm{Rh})\right)^{-1}_{\rm{CSA}}
    = 
    \frac{2}{15} B_{0}^2 
    \gamma_\mathrm{Rh}^2 
   \Delta\sigma^2 \tau_c
\end{equation}

where the shielding anisotropy $\Delta\sigma$ is defined as follows~\cite{kowalewski_book_2018}:

\begin{align}\label{eq:Deltasigma}
\Delta\sigma &= \frac{3}{2}(\sigma_{ZZ}-\sigma_\mathrm{iso})
=
-\frac{3}{2}\delta^\mathrm{aniso}
\end{align}

Equation~\ref{eq:RhT1theory} implies that the quadratic field-dependent coefficient $a$ for the \Rh $T_1^{-1}$ relaxation rate constant is given by
\begin{equation}
    a = \frac{2}{15}  
    \gamma_\mathrm{Rh}^2 
    \Delta\sigma^2 \tau_c
\end{equation}
The experimental estimate of the quadratic coefficient $a=0.023\pm 0.001\mathrm{\,s^{-1}\,T^{-2}}$ may be combined with the correlation time estimate $\tau_c\simeq 24.5\pm 1.5\mathrm{\,ps}$
to obtain the following experimental estimate of the \Rh shielding anisotropy: $|\Delta\sigma|=9900 \pm 540\mathrm{\ ppm}$. 

This is a very large number. Although prior estimates of the \Rh CSA are scarce in the literature, CSA values for heavy spin-1/2 nuclei are sometimes of a similar magnitude \cite{dechter_195pt_1984,paluch_analysis_2019,sparks_platinum-195_1986,venkatesh_structure_2020,benn_heavy_1985,bagno_nmr_2010,perras_indirect_2017,venkatesh_enhancing_2018,venkatesh_molecular_2022,venkatesh_proton-detected_2021}, with closely related platinum (II) compounds displaying $^{195}$Pt CSA values  on the order of 10,000 ppm \cite{dechter_195pt_1984,sparks_platinum-195_1986,venkatesh_enhancing_2018,venkatesh_proton-detected_2021}. To our knowledge, the only other measurements of \Rh CSAs, in  very different Rh(III) compounds, were on the order of $\sim$500-1500 ppm \cite{adams_activation_1987,phillips_observation_2006}.
This dramatic range is also typical\cite{benn_heavy_1985,venkatesh_molecular_2022,venkatesh_structure_2020} for heavy spin-1/2 nuclei.

Using ORCA~\cite{neese_orca_2020,stoychev_self-consistent_2018,stoychev_efficient_2018}, \Rh shielding tensors \hl{were computed at the TPSSh/SARC-ZORA-TZVPP level of theory using implicit solvation (CPCM}\cite{barone_quantum_1998,garcia-rates_effect_2020} \hl{for THF), the zeroth-order regular approximation (ZORA)}\cite{bouten_relativistic_2000,van_wullen_molecular_1998} \hl{for the inclusion of relativistic effects, GIAOs, the RI approximation}\cite{stoychev_efficient_2018} \hl{, and the tau-dependent correction as suggested by Dobson}\cite{dobson_alternative_1993,reimann_importance_2015,schattenberg_effect_2021} (see Supporting Information). The result is summarised in table~\ref{tab:RhCSA}.

The calculated CSA is somewhat smaller than the experimental estimate. Underestimation of CSAs calculated using the ZORA method has been reported for other heavy spin-1/2 nuclei \cite{alkan_spin-orbit_2018,autschbach_accuracy_2004,malkin_absolute_2013}, \hl{where} better agreement might be obtained with higher-order four-component relativistic calculations \cite{autschbach_accuracy_2004} or by accounting for the relativistic breakdown of the relationship between spin-rotation and the paramagnetic contribution to the anisotropy \cite{malkin_absolute_2013}. 

\begin{table}[bth]
\begin{tabular}{|l|l|}
\hline
Method                                  & 
$|\Delta\sigma|$ /ppm \\ \hline
Calculated    & 7070           \\ \hline
Experimental estimate                     
    & \hl{9900}  $\pm$ 540
\\ \hline
\end{tabular}
\caption{Estimates of the \Rh shielding tensor anisotropy $\Delta\sigma$ of Rh formate, defined in equation~\ref{eq:Deltasigma}. The computational estimate is given 
by quantum chemical calculation using ORCA~\cite{neese_orca_2020}. The experimental estimate is from the analysis of field-dependent \Rh relaxation in solution, as described in this paper.
\label{tab:RhCSA}
}
\end{table}

The origin of the zero-field contribution $T_1^{-1}(0)$ to the \Rh relaxation rate constant is currently unknown. As discussed in the Supporting Information, the \Rh-\Rh and \Rh-\Proton dipole-dipole couplings are much too weak to account for this term.  In the literature, the low-field relaxation of heavy spin-1/2 nuclei is often attributed to a spin-rotation mechanism. However, to our knowledge, this conclusion has not been supported by any theoretical or computational studies.

The experimental estimate of the \Rh $T_2$ is much shorter than the estimate of $T_1$ under the same conditions ($T_2=0.181\pm0.001\mathrm{\ s}$ as against $T_1=0.483\pm0.002\mathrm{\,s}$, in a field of $9.4\mathrm{\,T}$. We tentatively attribute the short $T_2$ value to the modulation of the isotropic chemical shift by ligand exchange at the axial positions. Other decoherence mechanisms, such as diffusion in the presence of inhomogeneous magnetic fields, are expected to be too weak to account for the observed $\Ttwo$ value in this case. 

In \hl{conclusion}, this paper has demonstrated methodology for the indirect estimation of \Rh $T_1$ and $T_2$ values by magnetization transfer to and from \Proton nuclei using the DualPol pulse sequence. 
Field-dependent \Rh $T_1$ measurements indicate a very large chemical shift anisotropy for the \Rh sites in the rhodium formate paddlewheel complex. The field-independent contribution to the \Rh relaxation rate constant is not fully understood at the current time.

\hl{A limitation of the methodology described here is the prerequisite of a spin system with direct scalar couplings between $^{103}$Rh nuclei and a proton, which is not present in all rhodium complexes. This limitation may be addressed via the use of a relay nucleus, such as $^{13}$C at natural abundance.}\cite{calo_triple_2021,benn_indirect_1988,mcfarlene_studies_1976,weske_hcag_2015}

\begin{acknowledgments}
We acknowledge funding from the European Research Council (grant 786707-FunMagResBeacons), and EPSRC-UK (grants EP/P009980/1, EP/P030491/1, EP/V055593/1). M.L. acknowledges financial support by the Max-Planck-Gesellschaft and the Max-Planck-Institut für Kohlenforschung.
We thank Alexander \hl{A.} Auer for advice on quantum chemical calculations. We thank Professor Brian E. Mann for advice and historical insights on rhodium NMR. We thank Alexey Kiryutin for sharing his designs for the sample shuttle.
\end{acknowledgments}

\section*{Author Declarations}

\subsection*{Conflict of interest}

The authors have no conflicts to disclose.

\section*{Data Availability Statement}

The data that support the findings of this study are available from the corresponding author upon reasonable request.

\bibliography{References/RhFormate}

\end{document}


\include{References/RefSets}
\title{The $^{103}$Rh NMR Spectroscopy and Relaxometry of the Rhodium Formate Paddlewheel Complex}
\author[1]{Harry Harbor-Collins}
\author[1]{Mohamed Sabba}
\author[1]{Gamal Moustafa}
\author[1]{Bonifac Legrady}
\author[1]{Murari Soundararajan}
\author[2]{Markus Leutzsch}
\author[1,a]{Malcolm H. Levitt}
\affil[1]{School of Chemistry, University of Southampton, SO17 1BJ, UK}
\affil[2]{Max-Planck-Institut für Kohlenforschung, Kaiser-Wilhelm-Platz 1, Mülheim an der Ruhr, 45470, Germany}
\maketitle
\blfootnote{$^a$ Electronic mail: mhl@soton.ac.uk}
\newpage
\section{Rhodium formate $^{13}$C $T_1$}
\subsection{Pulse sequence}
\begin{figure}[h]
\hspace*{-0.68cm}
\centering
\includegraphics[trim={1cm 1.5cm 1.3cm 2cm},clip,width=1\columnwidth]{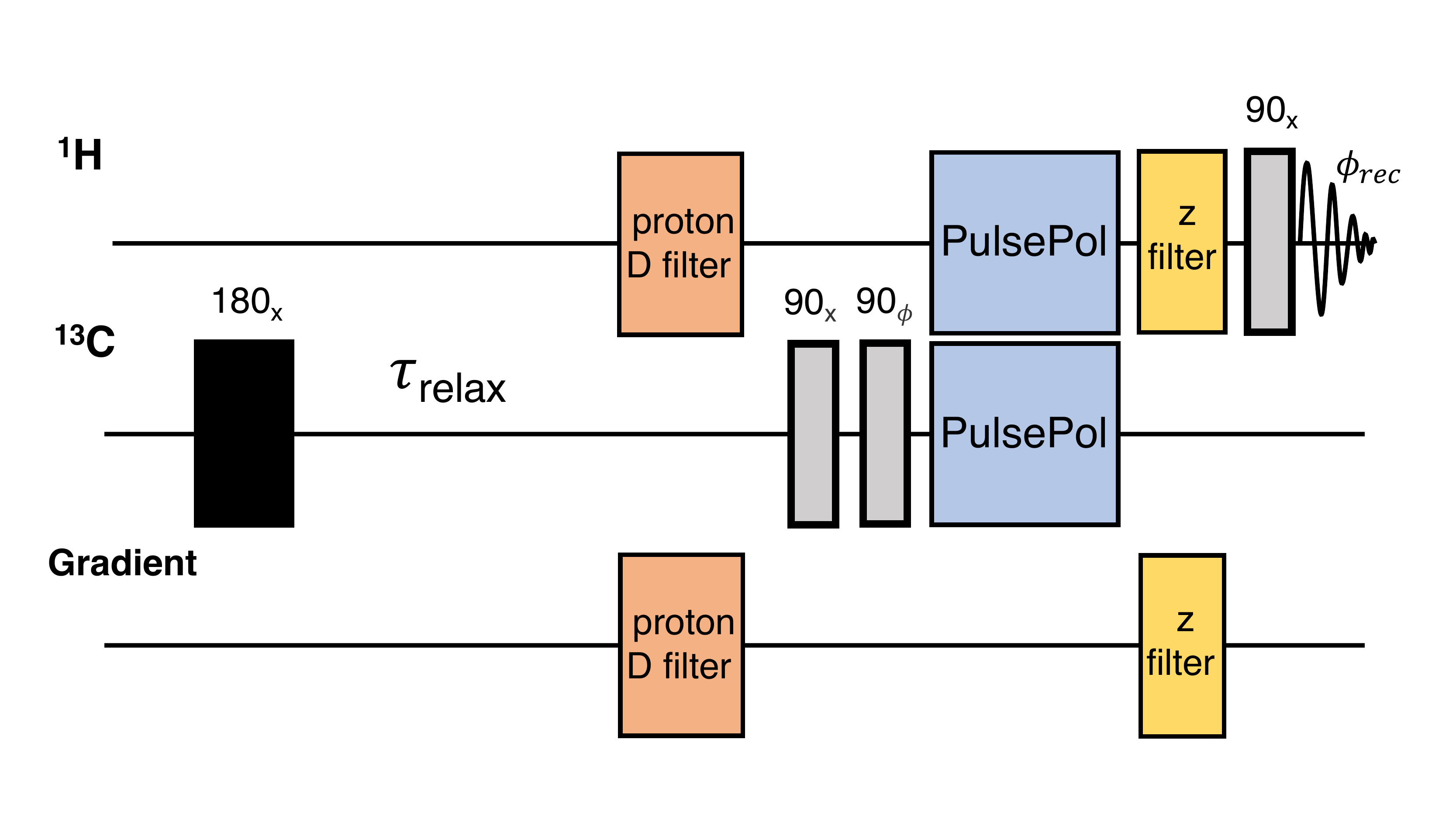}
\setlength{\belowcaptionskip}{-0pt}
\caption{\label{fig:SI_13CT1sequence} 
Sequence used for the indirect measurement of carbon T$_1$ through $^1$H NMR signals. Phase cycles are given by $\phi = [{-x,x}]$ and the receiver $\phi_{rec}=[x,-x]$. 
}
\end{figure}
$^{13}$C $T_1$ relaxation time constants were measured indirectly through $^1$H NMR signals using the sequence shown in figure~\ref{fig:SI_13CT1sequence}. Longitudinal $^{13}$C magnetisation is inverted by an initial 180$^{\circ}$ pulse, and allowed to relax towards equilibrium during the waiting interval $\tau_{relax}$.
A proton destruction filter is applied to eliminate any residual proton magnetisation generated during the waiting interval. Remaining $^{13}$C z-magnetisation, selected for by the two 90$^{\circ}$ pulses, is now transferred back to $^1$H z-magnetisation by a DualPol block and is selected for by a proton z-filter. A final$^1$H 90$^{\circ}$ pulse generates observable $^1$H transverse magnetization. The sequence is repeated with variation of the waiting interval $\tau_{relax}$ in order to follow the equilibration of longitudinal $^{13}$C magnetization.
\newpage
\subsection{Rhodium formate $^{13}$C $T_1$ decay curve}
\begin{figure}[h]
\hspace*{-0.68cm}
\centering
\includegraphics[trim={0cm 0cm 0cm 0cm},clip,width=0.85\columnwidth]{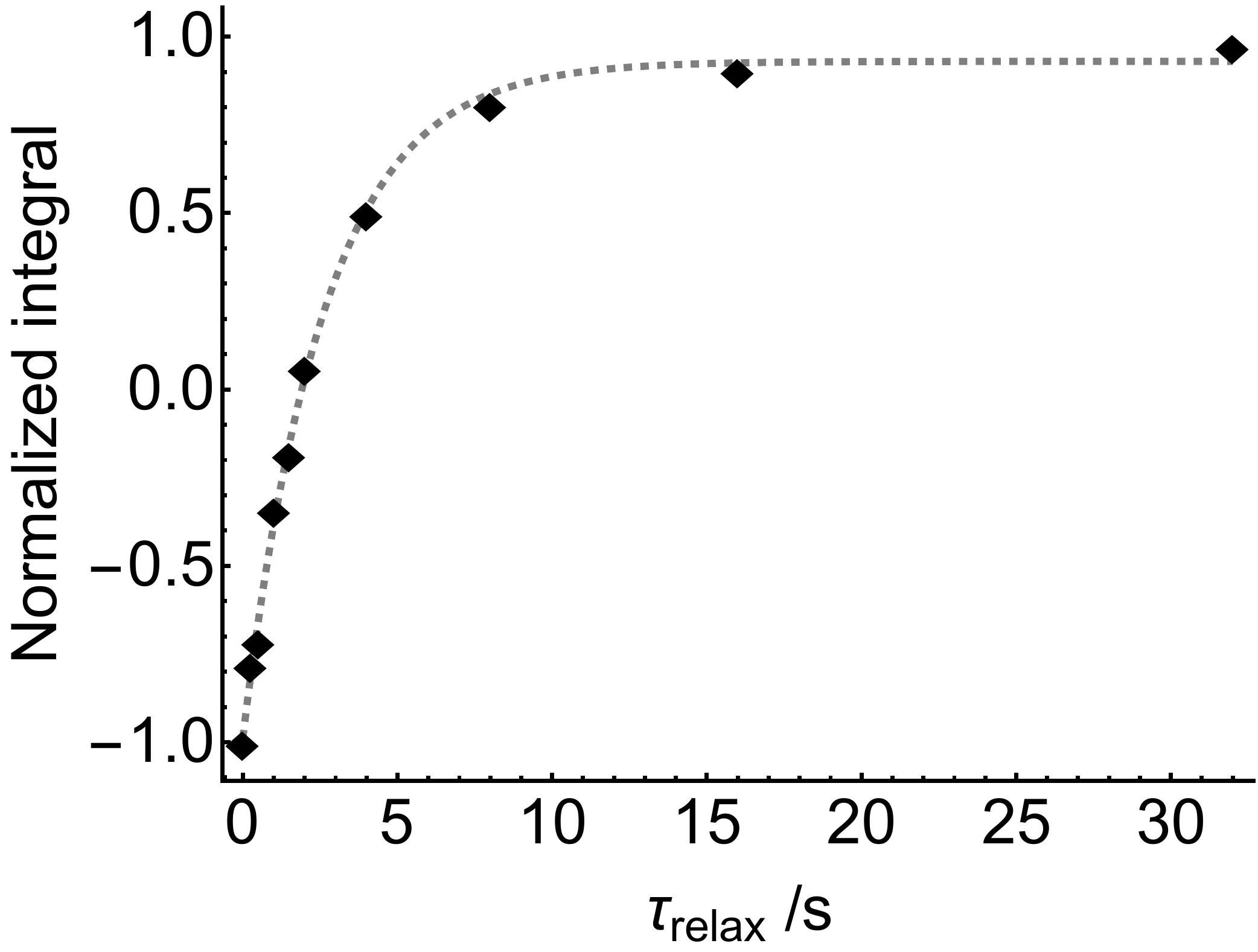}
\setlength{\belowcaptionskip}{-0pt}
\caption{\label{fig:SI_13CT1} 
Experimental $^{13}$C longitudinal magnetization recovery for rhodium formate dissolved in THF-d$_8$ at a field of 9.4~T. The integrals are normalised against the first data point.
}
\end{figure}
The trajectory of indirectly-detected $^{13}$C z-magnetization in a field of $9.4~\mathrm{T}$ is shown in figure~\ref{fig:SI_13CT1}. The trajectory fits well to a single-exponential decay with a time constant of $2.64\pm0.13$ seconds.
\newpage
\section{Quantum chemical calculations}
\subsection{Geometry optimisation}
\begin{figure}[h]
\hspace*{-0.68cm}
\centering
\includegraphics[trim={0cm 0cm 0cm 0cm},clip,width=0.85\columnwidth]{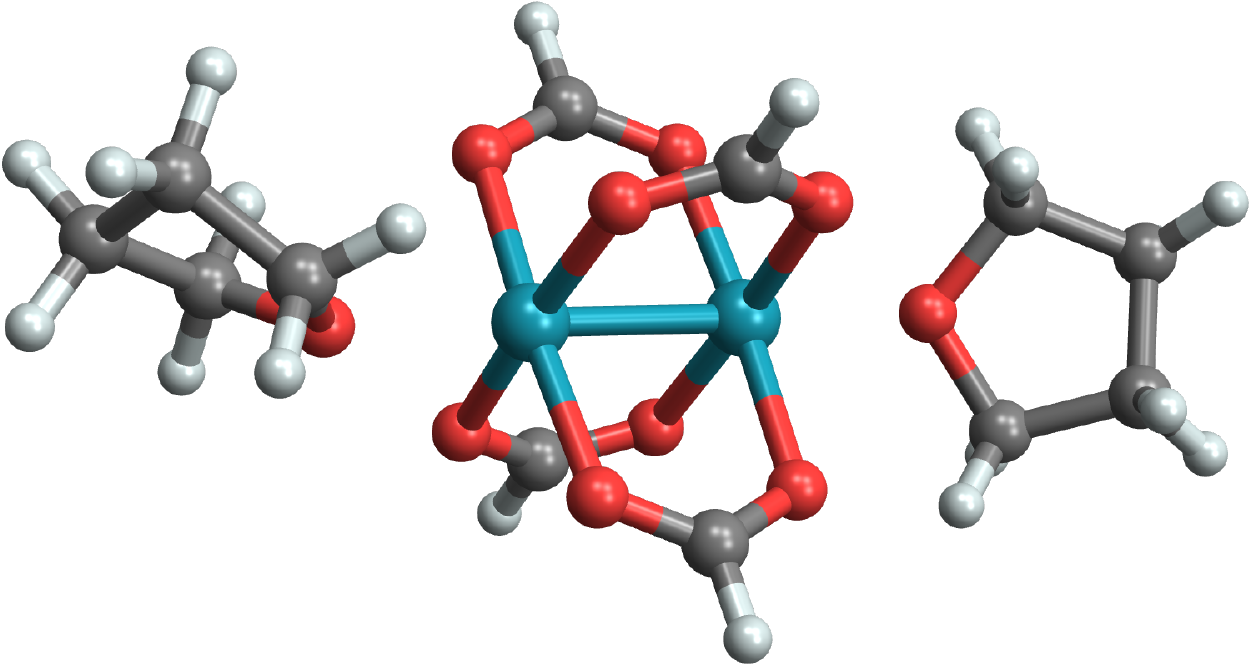}
\setlength{\belowcaptionskip}{-0pt}
\caption{\label{fig:SI_RhformateGeometry} 
Rhodium formate with axial ligation by THF molecules geometry optimised structure.
}
\end{figure}
The geometry of the rhodium formate complex axially ligated by solvent THF molecules was optimised using the ORCA program package version 5.0.3~\cite{neese_orca_2020}. Calculations were performed at the B3LYP/SARC-ZORA-TZVP level of theory. The .xyz file describing the optimized molecular coordinates is provided separately with the supplement.
\newpage
\subsection{Shielding tensor calculation}
The $^{103}$Rh shielding tensors were were computed at the TPSSh/SARC-ZORA-TZVPP level of theory using implicit solvation (CPCM\cite{barone_quantum_1998,garcia-rates_effect_2020} for THF), the zeroth-order regular approximation (ZORA)\cite{bouten_relativistic_2000,van_wullen_molecular_1998} for the inclusion of relativistic effects, GIAOs, the RI approximation\cite{stoychev_efficient_2018}  and the tau-dependent correction as suggested by Dobson\cite{dobson_alternative_1993,reimann_importance_2015,schattenberg_effect_2021}. The full ORCA input files and output files for the calculation are provided separately in the supplement.
\begin{figure}[h]
\hspace*{-1.5cm}
\centering
\includegraphics[scale=0.45]{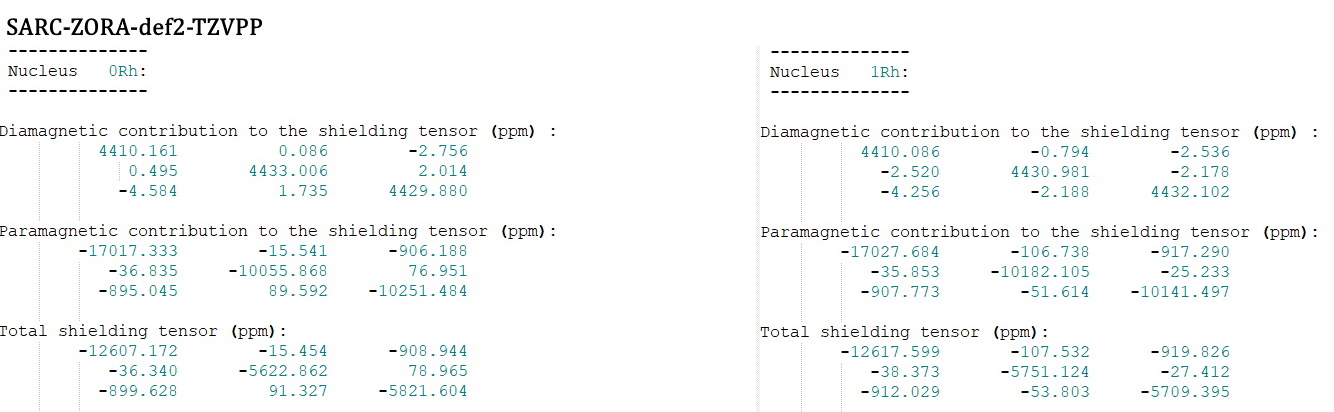}
\setlength{\belowcaptionskip}{-0pt}
\caption{\label{fig:SI_CSAtensors} 
Output of the CSA calculations in ORCA.
}
\end{figure}
\newpage
\subsection{Dipolar contributions to the $^{103}$Rh $(T_1)^{-1}$}
In the extreme narrowing limit, the contribution of a heteronuclear dipolar coupling to the $^{103}$Rh $(T_1)^{-1}$ relaxation rate constant (when the heteronucleus is decoupled) is given by,
\begin{equation}
(T_1(^{103}\rm{Rh}))_{RhX}^{-1} = 
   b_{RhX}^2 \tau_c
\end{equation}
Where $ b_{RhX}$ represents the dipolar coupling constant between the $^{103}$Rh and the heteronucleus $X$ and $\tau_c$ is the rotational correlation time.
In contrast, the contribution of the homonuclear $^{103}$Rh dipolar coupling to the $^{103}$Rh $(T_1)^{-1}$ relaxation rate constant is given by,
\begin{equation}
(T_1(^{103}\rm{Rh}))_{RhRh}^{-1} = 
   \frac{3}{2}b_{RhRh}^2 \tau_c
\end{equation}
Where $ b_{RhX}$ represents the dipolar coupling constant between the $^{103}$Rh nuclei and $\tau_c$ is again the rotational correlation time.
Using the optimised geometry in figure~\ref{fig:SI_RhformateGeometry}, the $^{103}$Rh-$^1$ H dipolar coupling was estimated to be $\sim2\pi$ 61~Hz; whereas the $^{103}$Rh-$^{103}$Rh homonuclear dipolar coupling constant was estimated to be $\sim2\pi$8~Hz. These are extremely small numbers - the $^{103}$Rh$(T_1)$ resulting purely from the dipolar mechanism is predicted to be on the order of a day. 
\section{Solid-state NMR}
\subsection{Solid-state $^{13}$C spectra}
Solid-state $^{13}$C spectra were straightforward to acquire at 14.1 T using a packed 4 mm rotor ($\sim$15 mg of sample). The pulse sequence used to acquire the solid-state $^{13}$C spectra that appear in the main text is illustrated below.

\begin{figure}[h]
\hspace*{-0.68cm}
\centering
\includegraphics[scale=1]{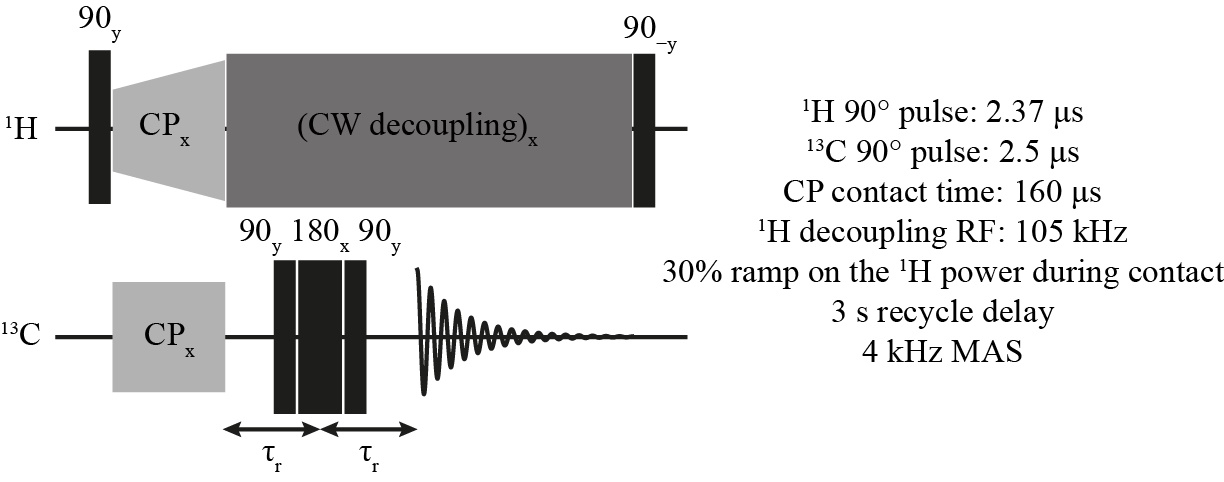}
\setlength{\belowcaptionskip}{-0pt}
\caption{\label{fig:SI_CP_parameters} 
Sequence used for the measurement of the $^{13}$C CPMAS spectra that appear in the main text. All of the relevant experimental parameters are illustrated in the figure.
}
\end{figure}
\bibliographystyle{unsrt}
\bibliography{References/RhFormate}